\documentclass[aps,prl,twocolumn,superscriptaddress,showpacs]{revtex4-1}
\usepackage{amsmath,amssymb}
\usepackage{graphicx}
\usepackage{xcolor}

\usepackage[final]{changes}
\reversemarginpar

\definechangesauthor[color=red]{AP}
\definechangesauthor[name = Georg, color=blue]{GK}
\definechangesauthor[color=green]{JF}
\definechangesauthor[name = JeanPascal, color=orange]{JP}
\definechangesauthor[name = JeanPascal New, color=magenta]{JPnew}

\begin{document}

\title{\added{Anomalous Anisotropy of the} Lower Critical Field and Meissner Effect in UTe$_2$}

\author{C. Paulsen}
\email{carley.paulsen@neel.cnrs.fr}
\affiliation{Univ. Grenoble Alpes, Institut N\'{e}el, C.N.R.S. BP 166, 38042 Grenoble France}
\author{G. Knebel}
\affiliation{Univ. Grenoble Alpes, CEA, Grenoble INP, IRIG, PHELIQS, F-38000 Grenoble, France}
\author{G. Lapertot}
 \affiliation{Univ. Grenoble Alpes, CEA, Grenoble INP, IRIG, PHELIQS, F-38000 Grenoble, France}
\author{D. Braithwaite}
\affiliation{Univ. Grenoble Alpes, CEA, Grenoble INP, IRIG, PHELIQS, F-38000 Grenoble, France}
\author{A. Pourret}
\affiliation{Univ. Grenoble Alpes, CEA, Grenoble INP, IRIG, PHELIQS, F-38000 Grenoble, France}
\author{D. Aoki}
\affiliation{Univ. Grenoble Alpes, CEA, Grenoble INP, IRIG, PHELIQS, F-38000 Grenoble, France}
\affiliation{Institute for Materials Research, Tohoku University, Ibaraki 311-1313, Japan}
\author{F. Hardy}
\affiliation{Institute for Solid-State Physics, Karlsruhe Institute of Technology, 76021 Karlsruhe, Germany}
\author{J. Flouquet}
\affiliation{Univ. Grenoble Alpes, CEA, Grenoble INP, IRIG, PHELIQS, F-38000 Grenoble, France} 
\author{J.-P. Brison}
\affiliation{Univ. Grenoble Alpes, CEA, Grenoble INP, IRIG, PHELIQS, F-38000 Grenoble, France}

%\abst{
\begin{abstract}
We report on low temperature susceptibility and magnetization measurements made on single crystals of the recently discovered heavy-fermion superconductor UTe$_2$  and compare the results with the two ambient pressure ferromagnetic superconductors URhGe and UCoGe.  Hysteresis curves in the superconducting phase show a familiar diamond shape superimposed on a large paramagnetic background.  The Meissner state was measured by zero field cooling in small fields of a few Oe as well as ac susceptibility measurements in small fields and resulted in 100\% shielding, with a sharp transition. However the field cooling Meissner-Ochsenfeld effect (expulsion of flux) was negligible in fields greater than just a few Oe, but becomes nearly 30\% of the perfect diamagnetic signal when the field was reduced to 0.01~Oe. The critical current due to flux pinning was studied by ac susceptibility techniques. Over the range in fields and temperature of this study, no signature of a ferromagnetic transition could be discerned. \added{The lower critical field $H_{\rm c1}$ has been measured along the three crystalographic axes, and surprisingly, the anisotropy of $H_{\rm c1}$ contradicts that of the upper critical field. We discuss this discrepancy and show that it may provide \replaced[id=GK]{additional}{ another} \deleted[id=GK]{strong} support for a magnetic field-dependent pairing mediated by ferromagnetic fluctuations in UTe$_2$.}
\deleted{The lower critical field $H_{c1}$ was found to be approximately 14 Oe for the $a$ and $c$ axis and 24 Oe along the $b$ axis. The upper critical field $H_{c2}$ along the $a$ axis was measured using bulk ac susceptibility measurements and found to be $H_{c2}= 5.4$ T in good agree with published resistivity measurements.}
\end{abstract}
%}

%%% Keywords are not needed any longer. %%%
%%%\kword{keyword1, keyword2, keyword3, \ldots}
%%%

%\begin{document}
\maketitle

%\section{Introduction}
Spin triplet superconductivity (SC) in itinerant ferromagnets close to the ferromagnetic (FM) -- paramagnetic (PM) instability was proposed four decades ago \cite{Fay1980}. The discovery of the coexistence of ferromagnetism and SC in UGe$_2$ opened the ``rush'' to a large variety of experiments \cite{Saxena2000}. The strong first order 
nature of the FM -- PM transition under pressure at $p_{\rm c} \approx 1.6$~GPa leads to SC occuring only in the FM domain in the pressure range from 1.2~GPa to 1.6~GPa; the maximum of the superconducting temperature $T_{\rm c}$ is 0.8~K, but the Curie temperature $T_{\rm Curie}  \approx 30$~K.\cite{Aoki2019a}

The field was enriched by the discoveries of two ambient pressure superconducting ferromagnets, URhGe \cite{Aoki2001}
and UCoGe \cite{Huy2007}, with $T_{\rm c} = 0.25$~K and 0.8~K, much lower than the respective $T_{\rm Curie} =9.5$~K and 2.7~K. The rapid suppression of $T_{\rm Curie}$ in UCoGe with pressure leads a the PM ground state above 1~GPa with the persistence of SC far above the critical pressure \cite{Bastien2016}. For both systems, the weakness of the FM interaction means that transverse magnetic fields ($H$) applied along the  $b$ axis, perpendicular to the easy \added{magnetization} axis $c$, of these orthorhombic crystals gives rise to a spectacular field-enhancement of SC \cite{Levy2005, Aoki2009, Aoki2019a}. 

The recent observation of SC in orthorhombic UTe$_2$ \cite{Ran2019, Aoki2019} at $T_{c} = 1.6$~K opens the possibility to study at ambient pressure spin-triplet SC in a system with a PM ground state located very close to a PM - FM instability. UTe$_2$ has the highest susceptibility \cite{Ikeda2006} and strong magnetic fluctuations \cite{TokunagaJPSJ2019} along the $a$ axis. However, the transverse field configuration with $H \parallel b$ attracted the most attention, due to the observation of a strong field-induced reinforcement of SC on approaching the metamagnetic field $H_{\rm m} \approx 35$~T.\cite{Miyake2019, Knafo2019, Knebel2019, Ran2019a}  Most of the published magnetization data in FM SC investigate the field dependence of the FM interaction by longitudinal or transverse field variation. \cite{Hardy2011a, Nakamura2017, Wu2017}. 

In URhGe and UCoGe the respective FM sublattice magnetization $M_0 = 0.4$~$\mu_{\rm B}$ and 0.07 ~$\mu_{\rm B}$ per U-atom produces an internal field of 800~G and 100~G far higher than the estimated value of the lower superconducting critical field $H_{\rm c1}$ of a few gauss. Thus even at $H =0$, self-induced vortices should occur, as shown for example in the magnetization studies on UCoGe \cite{Paulsen2012}. 

In this Letter we report  low temperature susceptibility and magnetization measurements on two crystals of UTe$_2$  ($T_{\rm c}=1.5$ and 1.6~K).
The experiments concentrate on (i) the \replaced{persistence}{surviving} of the PM state well below $T_{\rm c}$~K, (ii) the strength of the  \replaced{Meissner effect}{superconducting screening} in field-cooled (FC) experiments, (iii) the proof of a complete  \replaced{superconducting screening}{Meissner effect} in zero field cooled (ZFC) magnetization measurements, (iv) the determination of $H_{\rm c1}$, and (v) the determination of the London penetration depth \added{and of the} the superconducting coherence length from $H_{\rm c1}$ and \added{from} the upper critical field $H_{\rm c2}$.   We compare the results with the FM superconductors URhGe and UCoGe.\cite{suppl}  

All the measurements were made using two low-temperature superconducting quantum interference device (SQUID) magnetometers developed at the Institut N\'eel in Grenoble.  A unique feature of the setup is that absolute values of the magnetization can be measured using the extraction method  in a field range from 0.01~Oe up to 8~T (for details see the Supplemental Material \cite{suppl}). 

\begin{figure}[t]
\includegraphics[width=0.9\columnwidth]{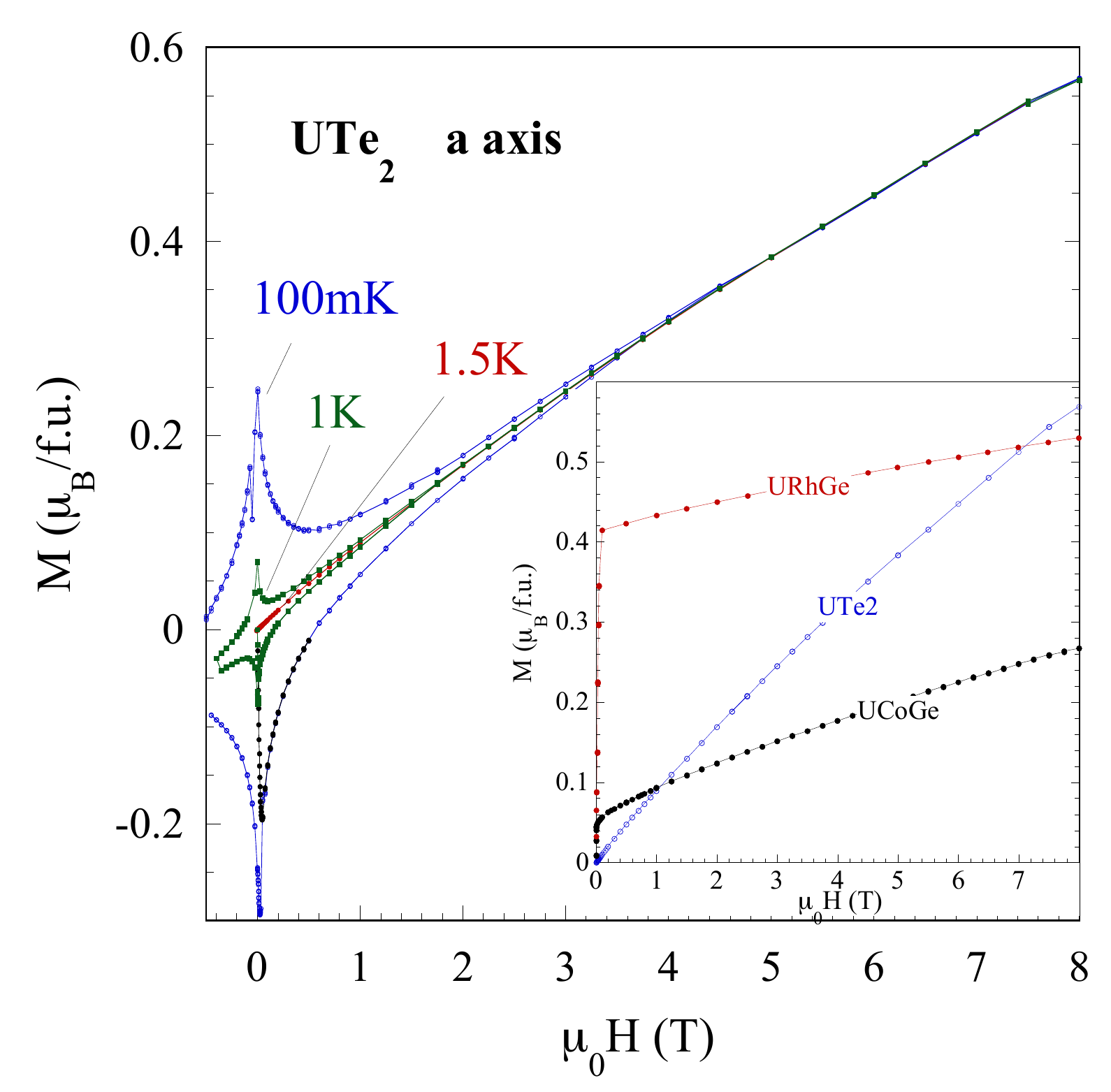}
\caption{Magnetization vs field for UTe$_2$ (sample 1) at 100~mK and 1~K below $T_{\rm c}$, and at 1.5~K just above $T_{\rm c}$ in the normal state with the field along the $a$ axis (easy axis). The insert shows the magnetization of UTe$_2$ (blue) at 1.5~K, URhGe (red) at 500~mK and UCoGe (black) at 600~mK, i.e.~just above their respective  $T_{\rm c}$ in the normal state, with the field along the easy axis. 
% (c) Hysteresis loops for UTe$_2$ at 1~K compared with UCoGe  at $T=70$~mK. (d) Minor hysteresis loops for UTe$_2$ measured at 800~mK, and UCoGe measured at 70~mK, on vastly different scales.
}
\label{Fig1}
\end{figure}

Figure \ref{Fig1} shows hysteresis loops for UTe$_2$ measured at 100~mK and 1~K below $T_{\rm c}$, and at 1.5K in the normal phase for UTe$_2$ with the field direction along the easy magnetization  $a$ axis. The slope of the initial magnetization  [\added{also shown later more clearly in Fig.~\ref{Fig4}(a)}] corresponds to 100\% shielding. The insert of Fig~\ref{Fig1} compares UTe$_2$ at 1.5K to the two FM superconductors URhGe and UCoGe, with the field applied along their easy $c$ axis,  at 500~mK and 600~mK above their respective superconducting states, but far below their respective Curie temperatures.  \replaced{The spontaneous moment of both ferromagnetic superconductors appears clearly, however, it is not a saturated moment: $M(H)$ keeps growing }{Note that both the later samples show a spontaneous moment $M(H)$ continues to grow}  with increasing field. Although there is no spontaneous moment for UTe$_2$,  the PM magnetization increases quickly and becomes larger than in UCoGe at about 1~T, and then greater than in URhGe above 7~T. 
%\deleted{The non linearity of $M (H)$ along the easy $c$ axis below $T_{\rm Curie}$ in UCoGe (decreasing $\frac{\partial M}{\partial H}$) points out that FM fluctuations decrease with increasing $H \parallel c$.} The \deleted{weak} susceptibility is weakest in URhGe  and it is connected with the fact that URhGe is a rather strong ferromagnet ($T_{\rm Curie} = 9.5$~K).  

The hysteresis loops for UTe$_2$ have a familiar  superconducting diamond shape which is superimposed on a very large PM background response.  
% take this shit out, who cares ? dosnt add anything to paper, .....The sharp dip in the signal at approximately -0.1~T is a magnetic avalanche and was very reproducible. In fact (depending on field ramping rate  at low temperature) a series of flux jumps or avalanches could be observed during the magnetization process, which gradually disappear above 500~mK. 
More hysteresis loops taken close to $T_c$ are shown in Fig.~S5 in the Supplemental Material, as well as a comparison with UCoGe in Figs.~S6 and S7.
In contrast to UCoGe where the superconducting and FM signals are fused together with the FM response dominating \cite{Paulsen2012}, our measurements of UTe$_2$ over the full temperature and field range show that there is no hint of FM behavior down to 80~mK in agreement with muon spin rotation ($\mu$SR) experiments.\cite{Sundar2019}    

% this is in supplemental now .....This is in contrast to UCoGe where the superconducting and FM signals are fused together with the FM response dominating as shown in Fig.~\ref{Fig1}(c) \cite{Paulsen2012}. In fact at first glance UCoGe does not look superconducting at all. However, the diamond shape response for UCoGe can be revealed by subtracting the hysteresis measured just above the superconducting transition from the low temperature data \cite{Deguchi2010}. Another way to see the pure superconducting response for UCoGe is to measure minor hysteresis loops using fields smaller than the FM coercive field, as shown in Fig.~\ref{Fig1}(d). Also shown  is a minor loop for UTe$_2$ but on vastly different scale. The form of these curves can be nicely fit using the Bean critical state model \cite{Bean1964}. 
%
\begin{figure}[t]
\includegraphics[width=0.9\columnwidth]{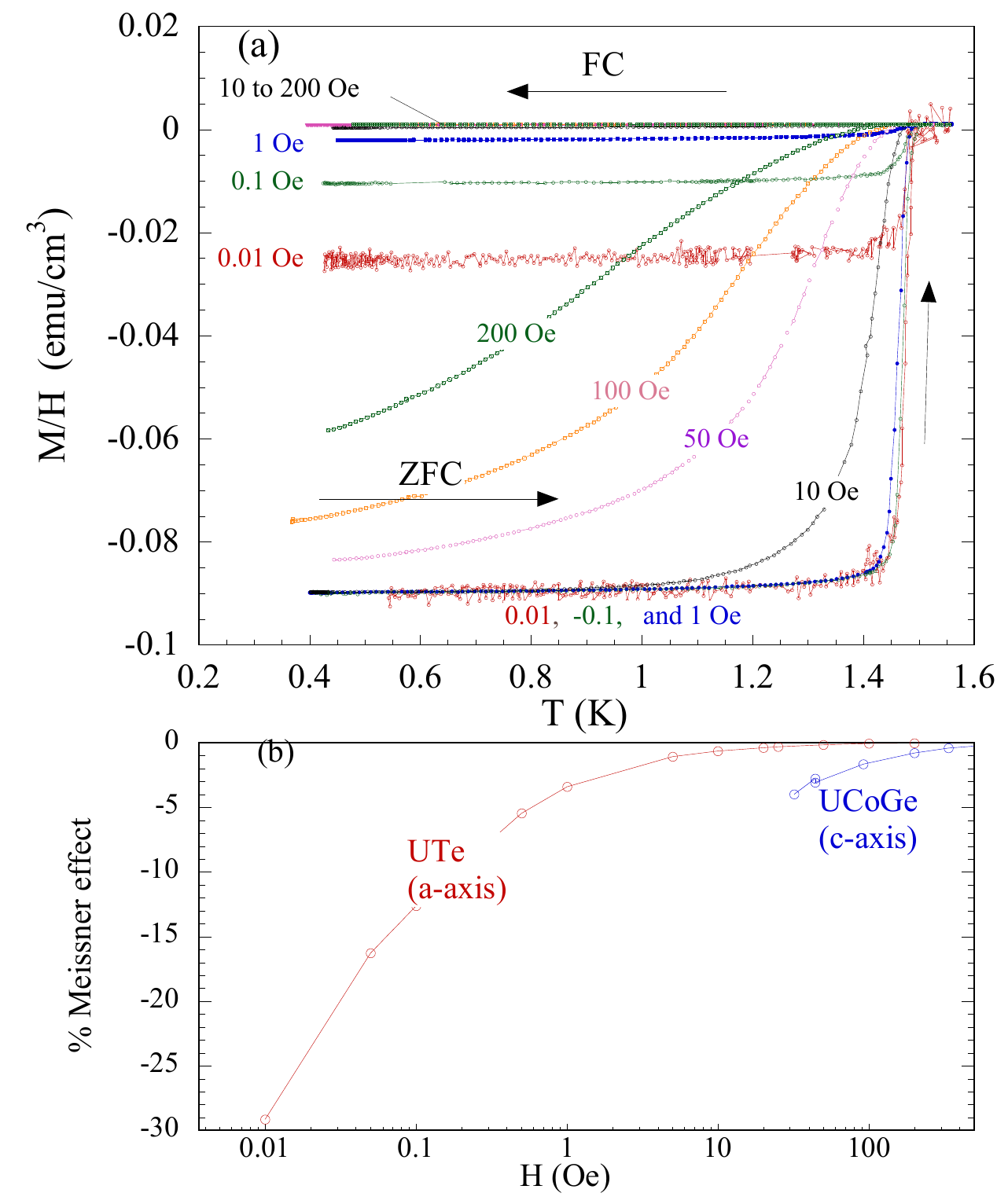}
\caption{(a) $M/H$ vs $T$ of UTe$_2$ (sample 1) for various applied fields ranging from 0.01 to 200 Oe from zero-field-cooled (ZFC) and field-cooled (FC) measurements. (b) The percent Meissner-Ochsenfeld effect (expulsion of flux) plotted against the applied field for UTe$_2$ (red points) and for UCoGe (blue points). Note that the field range in UCoGe cannot extend below 100mT as the sample needs to be mono-domain. Over the comparable field range, the expulsion of flux is much greater in UCoGe.}
\label{Fig2}
\end{figure}
% ￼￼

In Fig.~\ref{Fig2}(a) the dc susceptibility $M/H$ is plotted against temperature for various applied fields ranging from 0.01 to 200 Oe. Each curve was made by first zero-field-cooling (ZFC) the sample. A dc field was then applied and the sample was slowly warmed above $T_{\rm c}$, after which it was re-cooled in the same field, giving the field-cooled (FC) curve. In small dc fields the value of the ZFC susceptibility corresponds to 100\% shielding of the field (when demagnetization corrections are made), and the transition is sharp.  As the fields are increased, the transition becomes broader and shifts to lower temperatures. 
The FC susceptibility shows that the Meissner-Ochsenfeld effect (the reversible expulsion of flux as the sample is \added{field-}cooled and warmed through $T_{\rm c}$) for fields greater than a few Oe is negligible. However for very small fields, the effect becomes more important, reaching about 30\% expulsion in a field of 0.01~0e.

We compare this last result to UCoGe along the easy axis in Fig.~\ref{Fig2}(b). There are important differences. %in the measurements that need to be pointed out. 
First, the internal fields that are present in UCoGe, of the order 50-100 G, are much greater than $H_{\rm c1}$, and as a result UCoGe is always in the mixed state, and never achieves 100\% shielding. 
%% \deleted[id=JP,comment={I believe that there are other more important reasons for which UCoGe transitions are so broad} ]   {Thus the superconducting transition at $T_{\rm c}=0.5$~K is not sharp. } %\todo[inline]{I believe that there are other more important reasons for which UCoGe transitions are so broad} 
In addition, to measure the Meissner-Ochsenfeld effect in UCoGe means taking into account hysteresis and  a coercive field such that the applied field has no meaning while the sample is multi-domain \cite{Paulsen2012}. Nevertheless, a typical value of the percent expulsion of the flux from UCoGe compared to its effective shielding would be about 3\% expulsion at 50 Oe, which decreases with increasing field.  Although small, this is much greater than the Meissner-Ochsenfeld effect observed in UTe$_2$ in the same field range as can be seen in  Fig.~\ref{Fig2}(b). The other remarkable feature observed in Fig.~\ref{Fig2}(b) is the non-saturating rate of increase of the Meissner effect down to fields as low as 0.01~Oe in UTe$_2$: if any internal field exists due to a weak FM phase inside the superconducting phase, the resulting dipolar field has to be much smaller than 0.01~G, or in other words, the ordered moment \replaced{should be much }{is} smaller than $7\cdot 10^{-6}$~$\mu_B$.

\begin{figure}[t]
\includegraphics[width=0.9\columnwidth]{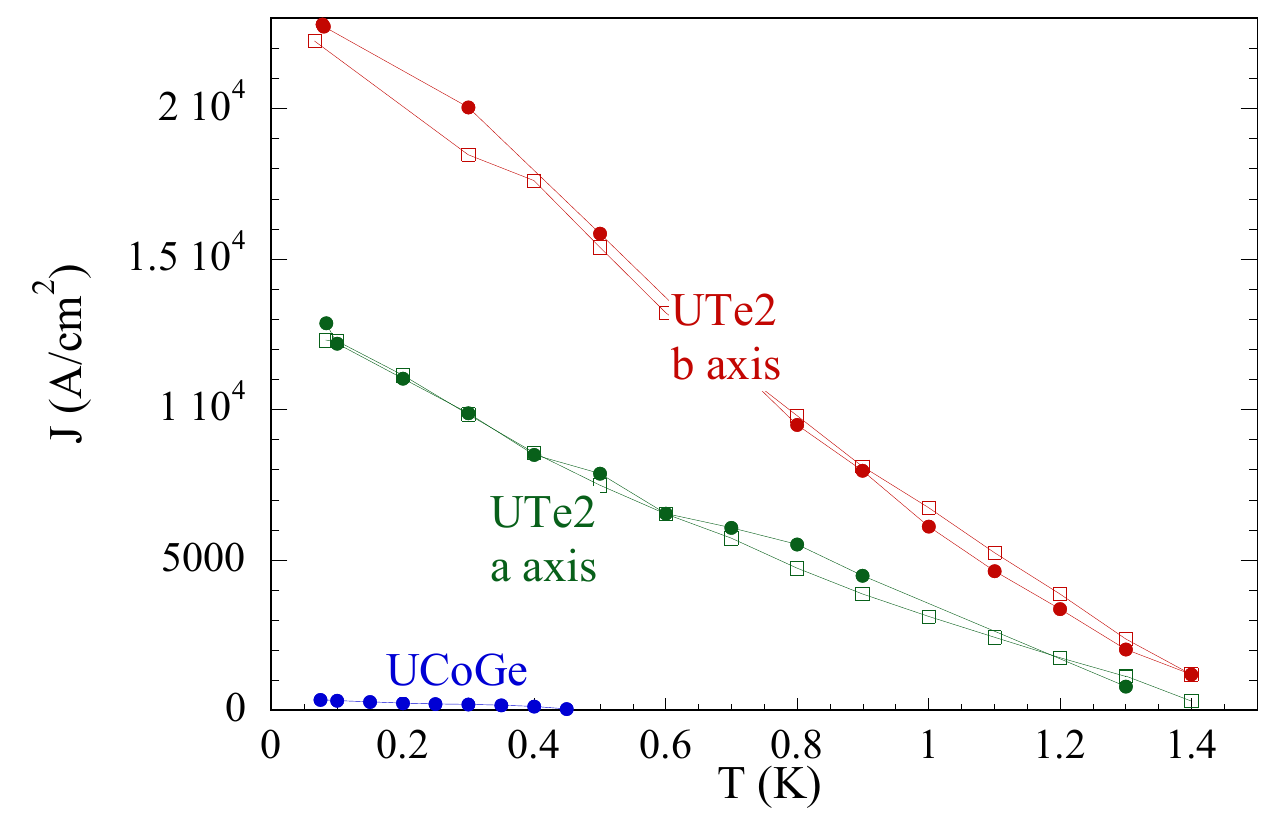}
\caption{Current density $J_{\rm c}$  derived from the critical state model vs temperature for UTe$_2$ (sample 2) measured along the $a$ and $b$ axes, and for UCoGe along the $c$ axis. The solid round points are data derived from the imaginary part of ac susceptibility, and the open square points are data derived from the remanent magnetization (discussed in the Supplemental Material). }
\label{Fig3}
\end{figure}

\begin{figure}[t]
\includegraphics[width=0.75\columnwidth]{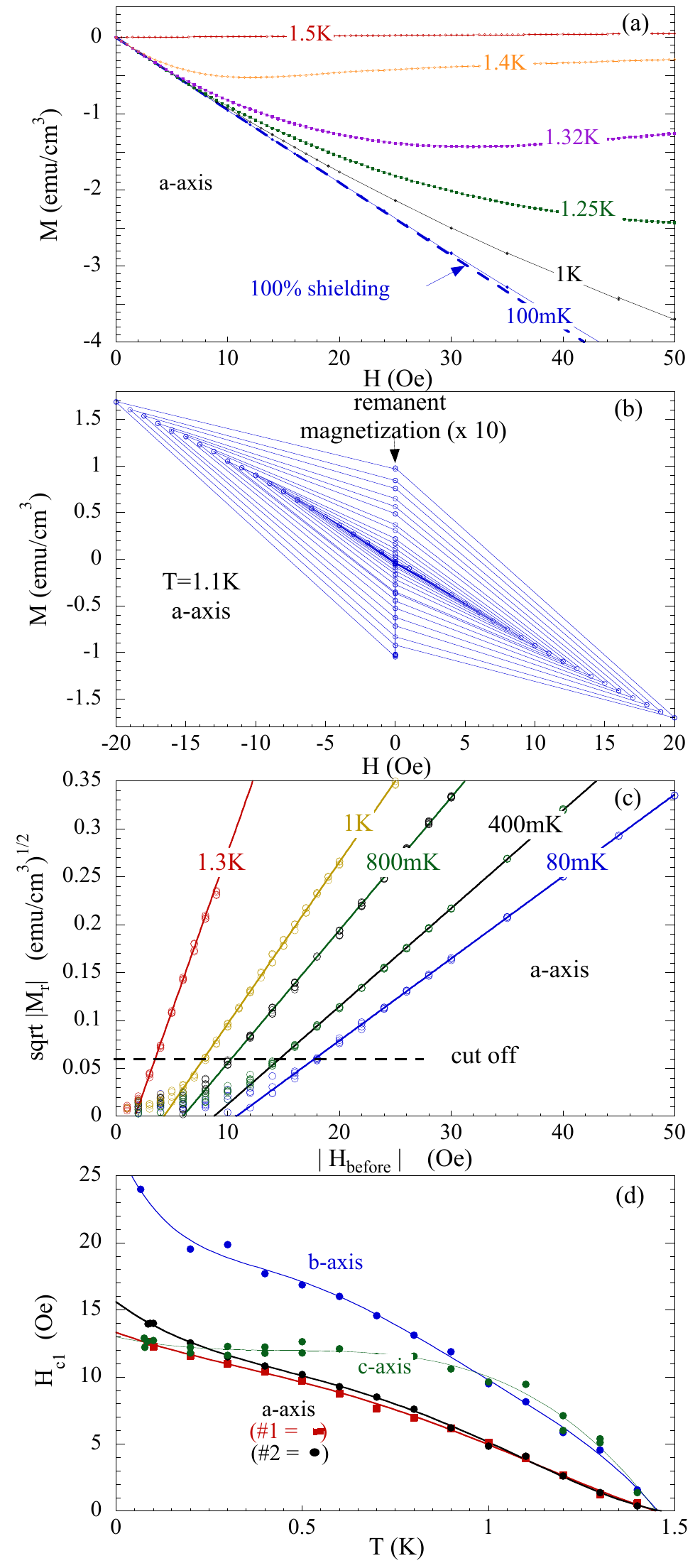}
\caption{(a) Initial magnetization $M$ vs $H$ taken a various constant temperatures for UTe$_2$ (sample 1) with the field along the $a$ axis. The dashed line is a linear fit to the 100~mK data at low fields, and represents 100\% shielding.
(b) Series of minor hysteresis loops for sample 1 at 1.1~K where the magnitude of the field was systematically increased in steps of 1 Oe, and then returned to zero to measure the point at which a remanent magnetization begins to appear in the sample (the remanent magnetization has been multiplied by 10).
(c) The remanent magnetizations is plotted as $\sqrt{M_{\rm r}}$ vs $H_{\rm before}$ for the $a$ axis for sample 1. The solid lines are linear fits to the data above the cut-off line.
(d) $H_{\rm c1} (T)$  for UTe$_2$ along the $a$ axis for sample 1 (red) and for sample 2 along the $a$ (black), $b$ (blue) and  $c$ (green) axis after correcting for demagnetization effects. The lines are guides to the eye. %The dashed lines are fits to data with the form $1- (T/T_{\rm c})^2$.
}
\label{Fig4}
\end{figure}

A strong hysteresis and a weak Meissner-Ochsenfeld effect \replaced{suggests }{implies} strong flux pinning. To confirm this we measured the ac susceptibility $\chi$ as a function of the ac driving field: An example can be seen in the Supplemental Material Fig.~S20  for the $b$ axis.  When flux begins to enter the sample, $\chi'$ and $\chi''$ of the ac susceptibility will deviate from their 100\% shielding values. The deviations are linear in the applied driving field and the slopes are proportional to $2/(J_{\rm c}D)$ for $\chi'$, and $2/(3\pi J_{\rm c} D)$ for $\chi''$, where $J_{\rm c}$ is the current density in the critical state model, and $D$ is the sample width, where we approximate the sample shapes as slabs\cite{Gomory1997}. The resulting $J_{\rm c} (H)$ for UTe$_2$ is plotted  in Fig.~\ref{Fig3}, along with $J_{\rm c}$ for UCoGe measured along the $c$ axis. Clearly flux pinning is far greater in UTe$_2$. 

The initial magnetization $M$ vs $H$ taken at various constant temperatures is shown in  Fig.~\ref{Fig4}(a). For each curve, the sample was first ZFC. The blue dashed line is a linear fit to the 100~mK data over a field range 0 - 10~Oe The slope of this fit (when corrected for demagnetization effects) corresponds to a susceptibility of $-1/4\pi$ ($-1$ in SI units) or 100\% shielding of the magnetic field. For a given temperature, as the field is increased, the curves deviate from this slope, and this is an indication that flux is entering the sample because $H_{\rm c1}$ has been exceeded. \added{As can be seen in Fig.~\ref{Fig4}(a), due to the strong pinning, flux enters the sample almost asymptotically and makes the determination of $H_{\rm c1}$ difficult. In addition, because the samples used in this study are not perfect ellipsoids, the field  enters the samples sooner around the edges due to demagnetization effects, further obscuring the real $H_{\rm c1}$.}

\added{A better way to determine $H_{\rm c1}$ is to measure the remanent magnetization as shown in Fig.~\ref{Fig4}(b) and 4(c) and discussed in detail in the the Supplemental Material. Figure~\ref{Fig4}(b) is an example of a series of minor hysteresis loops at 1.1~K for sample 1 where the magnitude of the field was systematically increased in small steps, and then returned to zero to measure the point where flux begins to enter the sample.  While  $H< H_{\rm c1}(T)$ the cycles are reversible. However, when $H_{\rm c1}(T)$ is exceeded, flux begins to enter the sample and the magnetization deviates from the $100\% $ shielding. When the field is then returned to zero, half the flux remains trapped in the sample, and a remanent moment $M_{\rm r}$ appears.  According to the Bean critical state model  $M_{\rm r} \propto H_{\rm before}^2$, i.e. the last field value before the field was reduced to zero and the remanence was measured \cite{Bean1964}. In order to estimate $H_{\rm c1}$, we modify the critical state model \replaced{following \cite{NaitoPRB1990} by assuming }{by assume} two regimes:  when  $H_{\rm before} < H_{\rm c1}$,  $M_{\rm r}=0$ and when $H_{\rm before}>H_{\rm c1}$,   $M_{\rm r}\propto (H_{\rm before}-H_{\rm c1})^2$.}

%According to the Bean critical state model for a slab shaped sample, the deviation from perfect screening is $\propto H^2$ : $-4\pi$M=H-$H^2/2H^*$ and it retains this flux when the field is removed, giving rise to a
% \highlight[id=JP, comment={Carley, with H=0 then $M_{\rm r}=0$ ??? should we put $H=H*$ ?}]{
%remanent magnetization $4\pi M_{\rm r}=H^2/4H^*$}, where $H^*$=$\pi J_cD/5$  using practical units H in Oe, D the thickness of the slab in cm, and  $J_c$ the critical current in amp/$cm^2$.

\added{In Fig.~\ref{Fig4}(c) we plot an example of the remanent magnetizations for sample 1 as $\sqrt{M_{\rm r}}$ vs $H_{\rm before}$. The solid lines are linear fits to the high field data above the cut-off line, thus omitting the low field values where we expect rounding to occur. The intercept gives $H_{\rm c1}$.}
 
\added{In Fig.~\ref{Fig4}(d)  $H_{\rm c1}$ along the $a$ axis  is plotted versus temperature for sample 1,  and for sample 2 for the $a$, $b$ and $c$ axes, where $H_{\rm c1}$ has been corrected for demagnetization effects: Both samples have a platelet shape, with the field applied parallel to the platelet, except for the $c$ axis measurements which was  performed perpendicular to the surface. We estimate the value of $H_{\rm c1}$ at $T=0$ to be  14~Oe along the $a$ and $c$ axis, and 24~Oe along the $b$ axis.}

\deleted{Another way to determine $H_{\rm c1}$ is shown in the Supplemental Material Fig.~S3 where minor hysteresis loops were made with the magnitude of the field systematically increased in small steps, and then returning to zero to measure the point where flux begins to enter the sample.  While  $H< H_{\rm c1}(T)$ the cycles are reversible. However when $H_{\rm c1}(T)$ is exceeded, flux begins to enter the sample and the magnetization deviates from the 100\% shielding slope. When the field is returned to zero, flux is trapped in the sample, and a remanent moment appears.   $H_{\rm c1}$  found by this method is shown in Fig.~\ref{Fig4}(b) for the $a$ and $b$ axis where $H$ has been corrected for demagnetization effects (both samples have a platelet shape, with the field applied parallel to the platelet). The value of $H_{\rm c1}$ we estimate to be  23~Oe along the $a$ axis, and 42~Oe along the $b$ axis. }

We have also studied $H_{\rm c2}$ along the $a$ axis using bulk ac susceptibility as shown in Fig.~S21 of the Supplemental Material
and we found   $H_{\rm c2}(0) = 5.4$~T along the $a$ axis, in good agreement with published resistivity measurements \cite{Ran2019, Aoki2019}.

\added{However, a comparison of $H_{\rm c1}$ and $H_{\rm c2}$ data reveals a major inconsistency. Indeed, close to  $T_{\rm c}$ in the Ginzburg-Landau (GL) regime, the anisotropy of $H_{\rm c1}$ should be opposite to that of  $H_{\rm c2}$ as $H_{\rm c1}H_{\rm c2}=H_{\rm c}^2( \ln\kappa +0.49)$, where $H_{\rm c}$ is the (isotropic) thermodynamic critical field and $\kappa=\frac{\lambda}{\sqrt{2}\xi}$ the anisotropic GL parameter. Hence, neglecting the anisotropy of $\kappa$ due to the logarithm, with $H_{\rm c2}^b>H_{\rm c2}^a \approx H_{\rm c2}^c$, one expects $H_{\rm c1}^b<H_{\rm c1}^a \approx H_{\rm c1}^c$. In contrast, Fig.~\ref{Fig4}(c) shows that the relative anisotropies anticipated from $H_{\rm c2}$ \replaced{are wrong in all }{are different for the three} directions. Quantitatively, this is a major effect, as the measured $H_{\rm c1}$ anisotropy between $a$ and $(b,c)$ is a factor 3, whereas the $H_{\rm c2}$ anisotropy between $(a,c )$ and $b$ is between 3 and 5, depending on the measurements notably for $H \parallel b$ (see, e.g., Ref.~\onlinecite{Aoki2019}).}
\deleted{These results can be compared to the thermodynamic critical field $H_{\rm c}$ extracted from heat capacity measurements.  At $T=0$ we find that $H_{\rm c} =490$~Oe (see Fig. S1 in Supplemental Material %\cite{suppl}
). For the $a$ axis, with  $H_{\rm c2}=5.4$~T, we get a Ginzburg-Landau (GL) parameter $\kappa \approx 78$ ($H_{\rm c2} = H_{\rm c} \sqrt{2} \kappa$). We can then estimate a value of $H_{\rm c1}\approx 21$~Oe from   $H_{\rm c1} = \frac{H_{\rm c}}{\sqrt{2} \kappa} \left( \ln(\kappa) + 0.497 \right)$.\cite{Hu1972}. This is very close to the observed value of $\approx 23$~Oe. From the value of $H_{\rm c2}$, with weak-coupling formula, we get a coherence length for the field $H\parallel a$ of $\xi_0^a=\sqrt{\xi_0^b\xi_0^c} \approx 8.3$~nm, and hence from $\kappa =  \frac{\lambda_L}{\sqrt{2}\xi}$ a London penetration depth for $H\parallel a$ of $\lambda_{\rm L} = 920$~nm. Microwave experiments have estimated $\lambda_L^{ab}$ for $H\parallel c$ to $947-1126$nm depending on the samples \cite{Metz2019}. The very large value of $\lambda_L$ confirms again that SC in UTe$_2$ is governed by  heavy carriers: even with the "lowest limit" estimate of 0.5 charge carriers per uranium \cite{NiuArXiv2019}, an effective mass of $160~m_0$ ($m_0$, the free electron mass) is necessary to account for $\lambda_L\approx 900~$nm. Such a large value is not unusual among heavy fermion superconductors (see e.g. \cite{OkazakiJPSJ2010}) and among the largest found in the literature, comparable to those found in very small carrier systems \cite{CollignonPRB2017}.
}
\deleted{So for $H \parallel a$  there is a very satisfying consistency between the measured  $H_{\rm c1}$, its estimation from $H_{\rm c2}$ and $H_{\rm c}$, as well as with the measured value of $\lambda_L$.  However the result found for $H_{\rm c1}\parallel b$ is very puzzling: $H_{\rm c2}$ in this direction is at least a factor 2 higher than for $H\parallel a$ at low fields, and reaches 35~T (the metamagnetic field) due to the reinforcement of SC in this direction.}\cite{Knebel2019, Ran2019a} \deleted{So $\kappa$ should be much larger in this direction  (at least $\kappa \approx 170$), and so $H_{\rm c1}$ much smaller (of order $11$~Oe). On the contrary, the experimental value is almost twice as large for $H\parallel a$, four times the estimated value from $H_{\rm c}$ (there is yet no estimate of $\lambda_L$ for $H\parallel b$, to compare also with measured $\lambda_L$ values). }

\added{Such a large discrepancy calls for an explanation. Measurement errors of $H_{\rm c1}$ for $H \parallel b$ might come from stronger pinning in this direction. The critical current has been found indeed twice as large for $H \parallel a$, see Fig.~\ref{Fig3}(b), but it is very unlikely that it could explain a factor of 10 error between the two directions. The sample geometries are also similar in both cases, excluding an explanation through a bad estimation of the demagnetisation corrections.}

\added{So the next step is to question the estimate of $H_{\rm c1}$ from the relations with the GL parameter: for single band $s$-wave superconductors, these relations hold even in very anisotropic cases (see e.g.~\cite{CavaPRB2007}). However UTe$_2$  is most likely $p$-wave, multigap (like most other heavy-fermion superconductors) and topological \cite{Jiao2020}. 
The last feature, implying the existence of low energy surface states might influence pinning, but if it has any influence on the determination of $H_{\rm c1}$, it should also be reflected in the critical current measurements. More interestingly, the multigap character (or the nodal gap structure \cite{KoganPRB2019} for one-dimensional (1D) irreducible representations) has been shown, theoretically (e.g., Ref.~\onlinecite{KoganPRB2002}) and experimentally (e.g., Ref.~\onlinecite{FletcherPRL2005}) to induce strong deviations of the anisotropy of the critical fields from the estimations through the GL parameter: But this holds at low temperature, not close to $T_{\rm c}$.}
%Essentially, close to $T_{\rm c}$ the relations should always hold, due to the validity of the  GL theory in this regime, but strong deviations can be expected on cooling. So the most surprising is the contradiction between the anisotropy of  $H_{\rm c1}$ and $H_{\rm c2}$ close to $T_{\rm c}$. Our measurements (see Fig.\ref{Fig4}b) leave little doubt that the anisotropy of $H_{\rm c1}$ observed at low temperature is preserved close to $T_{\rm c}$. However, there is presently no robust picture of the $H_{\rm c2}$ anisotropy close to $T_{\rm c}$ in UTe$_2$  (see for example \cite{Aoki2019}). Moreover, the direct effect of field on the strength of the superconducting pairing mechanism can also affect in both ways the initial slope of $H_{\rm c2}$ \cite{Wu2017}. 
%%%%% Added sentence 0/03/2021 %%%%

\added[id=JPnew]{Last, several works have reported double transitions in their UTe$_2$ crystals at zero pressure \cite{ThomasSciAdv2020, HayesArXiv2020}. 
Note however that $H_{\rm c1}$ and $H_{\rm c2}$ measurements are only sensitive to the upper transition, until both transition eventually cross under field, which does not happen below 8~T according to Ref. \cite{HayesArXiv2020}. 
Hence, whatever the origin of this double transition (intrinsic or due to inhomogeneities), it cannot help to understand the reported anomalous relative anisotropies of  $H_{\rm c1}$ and  $H_{\rm c2}$.

A possible explanation \added[id=GK]{for the puzzling anisotropy of $H_{\rm c1}$}, which cannot be completely ruled out, is a low-field change of slope of $H_{\rm c2}$.
This is suggested for example by specific heat measurements \cite{KittakaPhysRevRes2020}, which find a weak anisotropy between $a$ and $b$ axis due to a strong curvature of $H_{\rm c2}$ along $a$ (hence a much larger slope at low fields), but still a smaller slope along the $c$ axis. 
This may partially help \added[id=GK]{to reduce the discrepancy between $a$ and $b$ anisotropies of $H_{\rm c1}$ and $H_{\rm c2}$, but is still not enough to explain the $H_{\rm c1}$ result, requiring $H_{\rm c2}^a \gg H_{\rm c2}^b \sim H_{\rm c2}^c$.}
}

%The only effect we could think of presently to explain this anisotropy issue is the fact 
\added{A key to understanding this anomalous anisotropy might be to take into account that
the pairing mechanism in UTe$_2$ can be tuned by a magnetic field \cite{Ran2019,Knebel2019}, a mechanism inducing a change of slopes of $H_{\rm c2}$ \cite{Nakamura2017, Wu2017}, which is not taken into account by GL. 
A field-dependent pairing induces a field dependence of the ``bare" critical temperature  $T_{\rm c}(H)$ that is independent of the mixed state formation, as well as of the Fermi velocities ($v_F$)\cite{Wu2017}. 
Hence, with field-dependent pairing, both critical fields are functions of field (through $T_{\rm c}$ and $v_F$) and temperature  and it is easy to show that the measured slope at $T_{\rm sc}$ is: $\frac{dH_{\rm ci}}{dT}=\frac{\left(\frac{\partial H_{\rm ci}}{\partial T}\right)_{\rm H}}{1+\frac{dT_{\rm sc}}{dH} \left(\frac{\partial H_{\rm ci}}{\partial T}\right)_{\rm H} }$. Here, $\left(\frac{\partial H_{\rm ci}}{\partial T}\right)_{\rm H}$ is the ``usual" slope defined for field-independent pairing, for which GL relations should hold. 
This relation between the measured $\frac{dH_{\rm ci}}{dT}$ and $\frac{\partial H_{\rm ci}}{\partial T}$ shows that, because $H_{\rm c1}$ is three to four orders of magnitudes smaller than $H_{\rm c2}$, corrections due to $\frac{dT_{\rm sc}}{dH}$ should be negligible on $H_{\rm c1}$, which should reflect the true bare anisotropies. 
A $\frac{dT_{\rm sc}}{dH} \approx -0.1$~T/K for $H \parallel a$ and $+0.1$~T/K for $H \parallel b$, (assuming $\frac{dT_{\rm sc}}{dH} \approx 0$ for $H \parallel c$) could reconcile the lower and upper critical field measurements (see Supplemental Material for more details \cite{suppl}).}

To summarize, the surprising contradiction between lower and upper critical field anisotropies can be understood as a manifestation of the strong field dependence of the pairing strength in UTe$_2$, which is strongly suppressed along the easy axis and boosted along the hard $b$ axis. 
Moreover, this is an indication that pairing is suppressed by field along the easy axis in this system, an effect comparable to, although weaker than in UCoGe \cite{Nakamura2017, Wu2017}, which can be seen as additional \deleted[id=GK]{very strong} support for ferromagnetic fluctuations mediated pairing.

Moreover, from our very low-field measurements of the Meissner state, we can put an upper limit to any FM ordered moment above 100~mK of $7\times 10^{-6}$~$\mu_B$ in UTe$_2$. 
Restricted Meissner-Ochsenfeld expulsion is coherent with the observed strong pinning. 
A possible link between the present strong pinning and singular topological properties of the superconducting phase deserve to be clarified. 
%Along the easy magnetization $a$ axis, excellent agreements are found between $H_{\rm c}$, $H_{\rm c1}$ and $H_{\rm c2}$. The derived values of $\lambda_L$ and $\xi_0$ agree with previous determinations, and are among the largest values found in the literature, pointing to very strong electronic correlations. A puzzle is that the anisotropy of $H_{\rm c1}$  along $a$ and $b$ axis is opposite to that expected from the anisotropy of $H_{\rm c2}$, even close to $T_{\rm c}$. New systematic experiments are required to elucidate this enigma, notably on $H_{\rm c2}$.

\begin{acknowledgments}
We thank K.~Behnia and T.~Klein for fruitful discussions. 
We acknowledge the financial support of the Cross-Disciplinary Program on Instrumentation and Detection of CEA, the French Alternative Energies and Atomic Energy Commission, and KAKENHI (JP15H05882, JP15H05884, JP15K21732, JP16H04006, JP15H05745, JP19H00646).
\end{acknowledgments}
%
%\bibliographystyle{jpsj}
%\bibliography{UTe2}
%
\bibliographystyle{apsrev4-1}	
\bibliography{UTe2Hc1}

\end{document}

\title{Supplemental Material for \\Anomalous Anisotropy of the Low Temperature Susceptibility and Magnetization measurements of UTe$_2$}

\author{C. Paulsen}
\affiliation{Univ. Grenoble Alpes, Institut N\'{e}el, C.N.R.S. BP 166, 38042 Grenoble France}
\author{G. Knebel}
\affiliation{Univ. Grenoble Alpes, CEA, Grenoble INP, IRIG, PHELIQS, F-38000 Grenoble, France}
\author{G. Lapertot}
\affiliation{Univ. Grenoble Alpes, CEA, Grenoble INP, IRIG, PHELIQS, F-38000 Grenoble, France}
\author{D. Braithwaite}
\affiliation{Univ. Grenoble Alpes, CEA, Grenoble INP, IRIG, PHELIQS, F-38000 Grenoble, France}
\author{A. Pourret}
\affiliation{Univ. Grenoble Alpes, CEA, Grenoble INP, IRIG, PHELIQS, F-38000 Grenoble, France}
\author{D. Aoki}
\affiliation{Univ. Grenoble Alpes, CEA, Grenoble INP, IRIG, PHELIQS, F-38000 Grenoble, France}
\affiliation{Institute for Materials Research, Tohoku University, Ibaraki 311-1313, Japan}
\author{F. Hardy}
\affiliation{Institute for Solid-State Physics, Karlsruhe Institute of Technology, 76021 Karlsruhe, Germany}
\author{J. Flouquet}
\affiliation{Univ. Grenoble Alpes, CEA, Grenoble INP, IRIG, PHELIQS, F-38000 Grenoble, France}
\author{J.-P. Brison}
\affiliation{Univ. Grenoble Alpes, CEA, Grenoble INP, IRIG, PHELIQS, F-38000 Grenoble, France}

\maketitle

In this Supplemental Material we show complementary data to those presented in the main article. \added[id=JP]{Furthermore, we give more details on the analysis of the anisotropies of $H_{\rm c1}$ and $H_{\rm c2}$}.

\section{Details on the sample growth}
The samples of UTe$_2$ were synthesized at the CEA-Pheliqs \replaced[id=JP]{laboratory}{Grenoble} using chemical vapor transport. The starting elements were 6N Te and pure depleted Uranium, and the transport agent was Iodine (5mg/cm$^3$), flowing from the source at 1060$^\circ$ toward cold end at 1000$^\circ$ over a period of 10 days. The high quality of the samples was checked by x-ray Laue patterns and SEM XR microanalysis. Sample 1 was studied along the $a$ axis (easy axis) and had a RRR= 16. Specific heat measurements on this sample show a very sharp superconducting transition at $T_c = 1.5$~K. Sample 2 was measured along the $a$ $b$ and $c$ axis and this sample has a $T_c= 1.6$~K. Details of the sample preparations for URhGe and UCoGe can be found elsewhere.\cite{Aoki2011}

\section{Details of the experimental setups, sample shapes and demagnetization corrections}
The measurements were made using two low temperature SQUID magnetometer\added[id=JP]{s} developed at the Institut N\'{e}el in Grenoble. Both magnetometers are equipped with a miniature dilution refrigerator capable of cooling samples below 80 mK. A unique feature of the setup is that absolute values of the magnetization can be \replaced[id=JP]{obtained}{made} by using the extraction method, without heating the sample. One magnetometer has an 8~T superconducting magnet. The second one has a smaller 0.4~T magnet and is dedicated for low fields.
The initial low field environment for the second magnetometer is made by using mu\added[id=JP]{-} metal shields outside the dewar to reduce\deleted[id=JP]{d} the earth's ambient field to below 10~mG. 
A superconducting lead shield inside the dewar traps and stabilized this field. An active shield place\added[id=JP]{d} just inside the superconducting shield and in series with main solenoid insure\replaced[id=JP]{s}{d} that the field is never above a few Gauss near the lead superconducting shield.

\begin{figure}[h]
\makeatletter
\renewcommand{\thefigure}{S\@arabic\c@figure}
\makeatother
\begin{center}
\includegraphics[width=0.8\columnwidth]{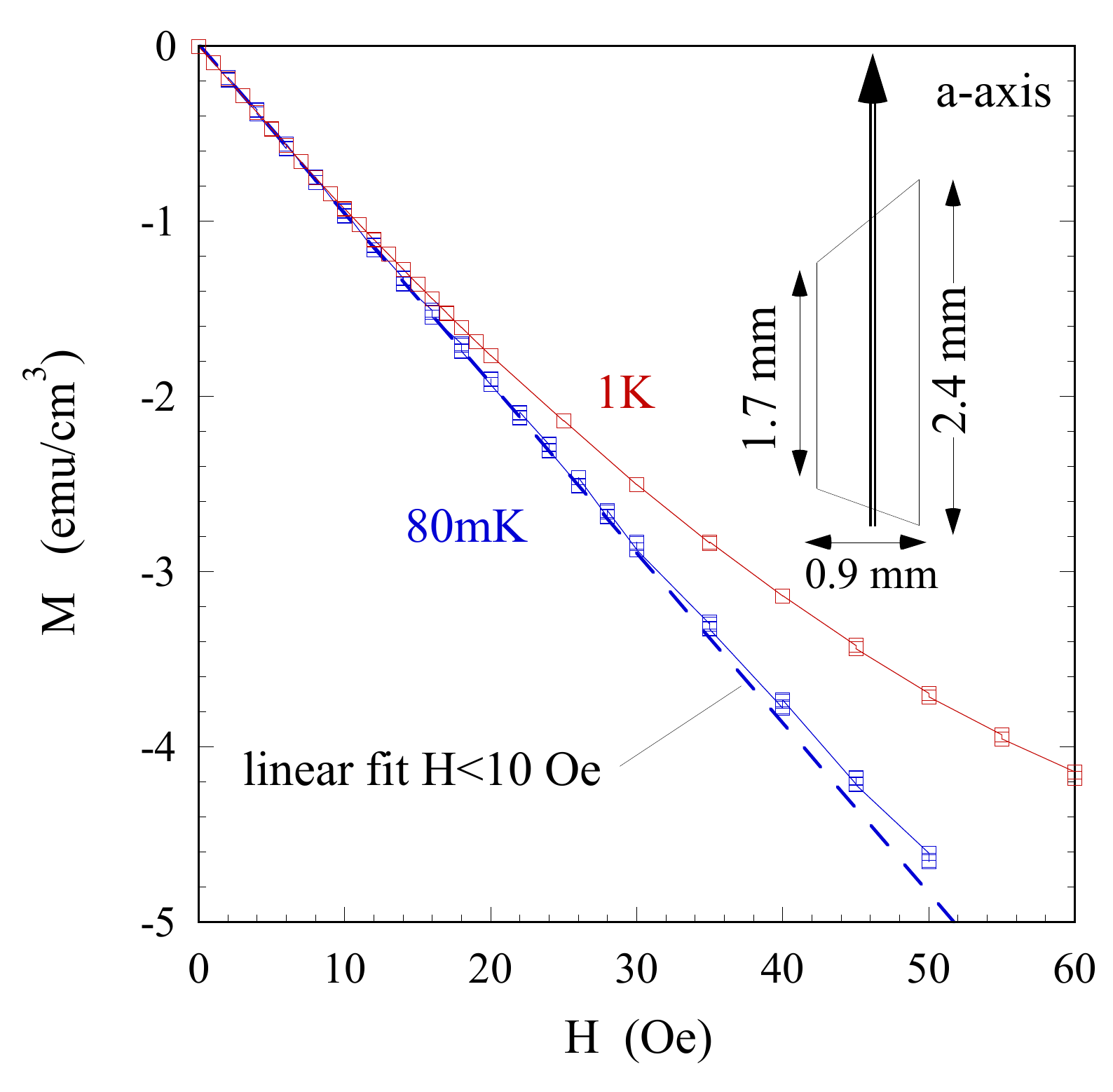} 
\caption{$M$ vs $H$ along the $a$ axis for sample 1, taken at 80~mK (blue) and 1~K (red). The dashed line is a fit to the 80~mK data for points less than 10 gauss, giving an applied susceptibility $M/ H_{\rm applied} =-0.094$ emu/cm$^3$. The rough shape of sample 1, and the direction of the applied field are shown, the sample thickness was approximately 0.4~mm. }

\label{suppl_fig1}
\end{center}
\end{figure}

\begin{figure}[h]
\makeatletter
\renewcommand{\thefigure}{S\@arabic\c@figure}
\makeatother
\begin{center}
\includegraphics[width=0.8\columnwidth]{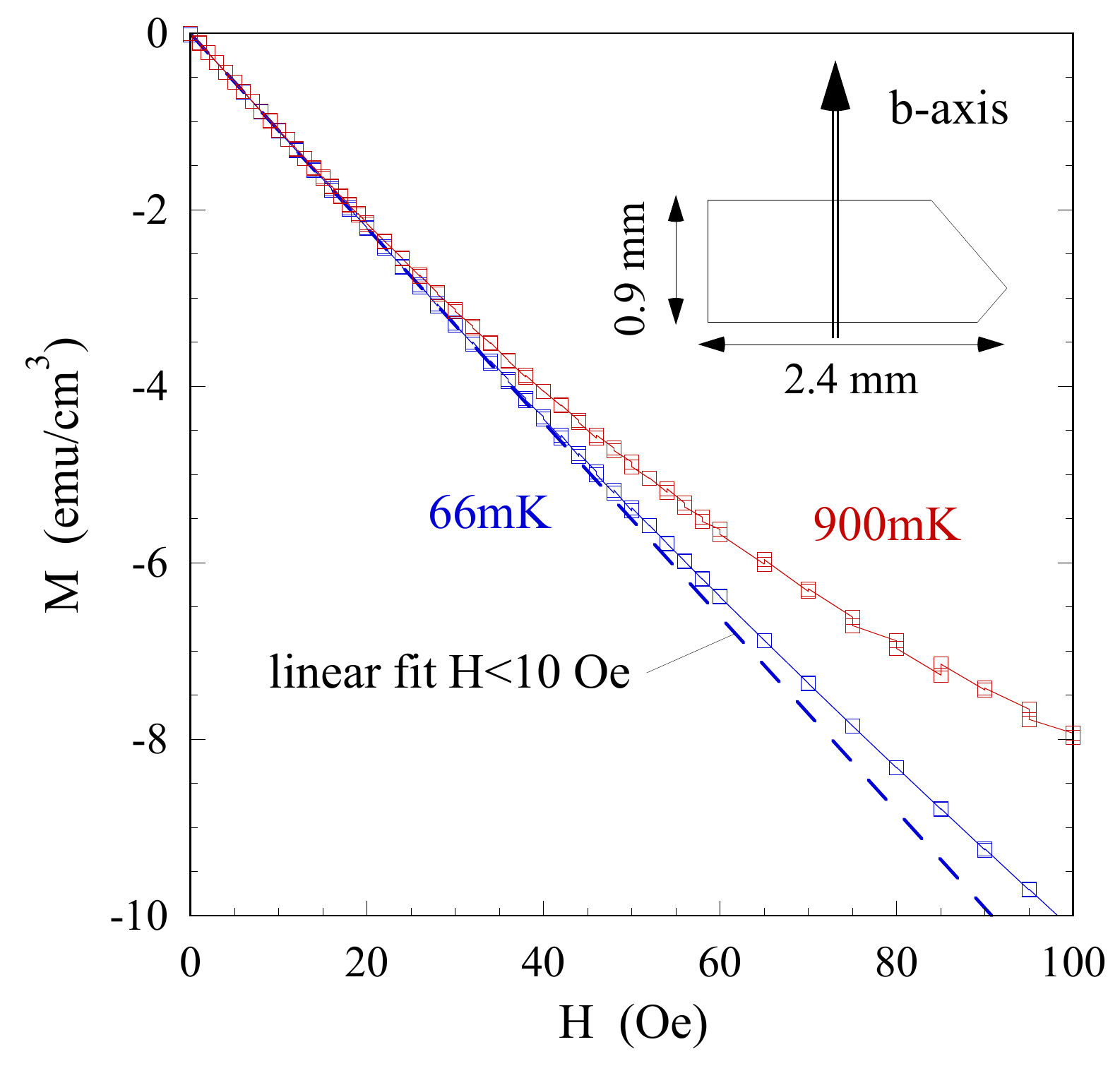} 
\caption{$M$ vs $H$ along the $b$ axis for sample 2, taken at 66~mK (blue) and 900~mK (red). The dashed line is a fit to the 66~mK data for points less than 10 gauss, giving an applied susceptibility $M/ H_{\rm applied} =-0.109$~emu/cm$^3$. The shape of sample 2 and the field direction are shown in the figure. The sample thickness was approximately 0.3~mm }
\label{suppl_fig2}
\end{center}
\end{figure}

Nevertheless when the applied field becomes greater than 100 Oe or so, flux starts to become trapped in the superconducting wire that makes up the solenoid and in the superconducting flux transformer itself, and this becomes more and more important the larger the field. To minimize these effects, after each field change, the flux transformer is pulse-heated to relieve the flux strain and insure uniform field around the sample. If needed, demagnetizing routines can reduce the trapped flux in the solenoid to below 100~mG. 

\begin{figure}[h]
\makeatletter
\renewcommand{\thefigure}{S\@arabic\c@figure}
\makeatother
\begin{center}
\includegraphics[width=0.8\columnwidth]{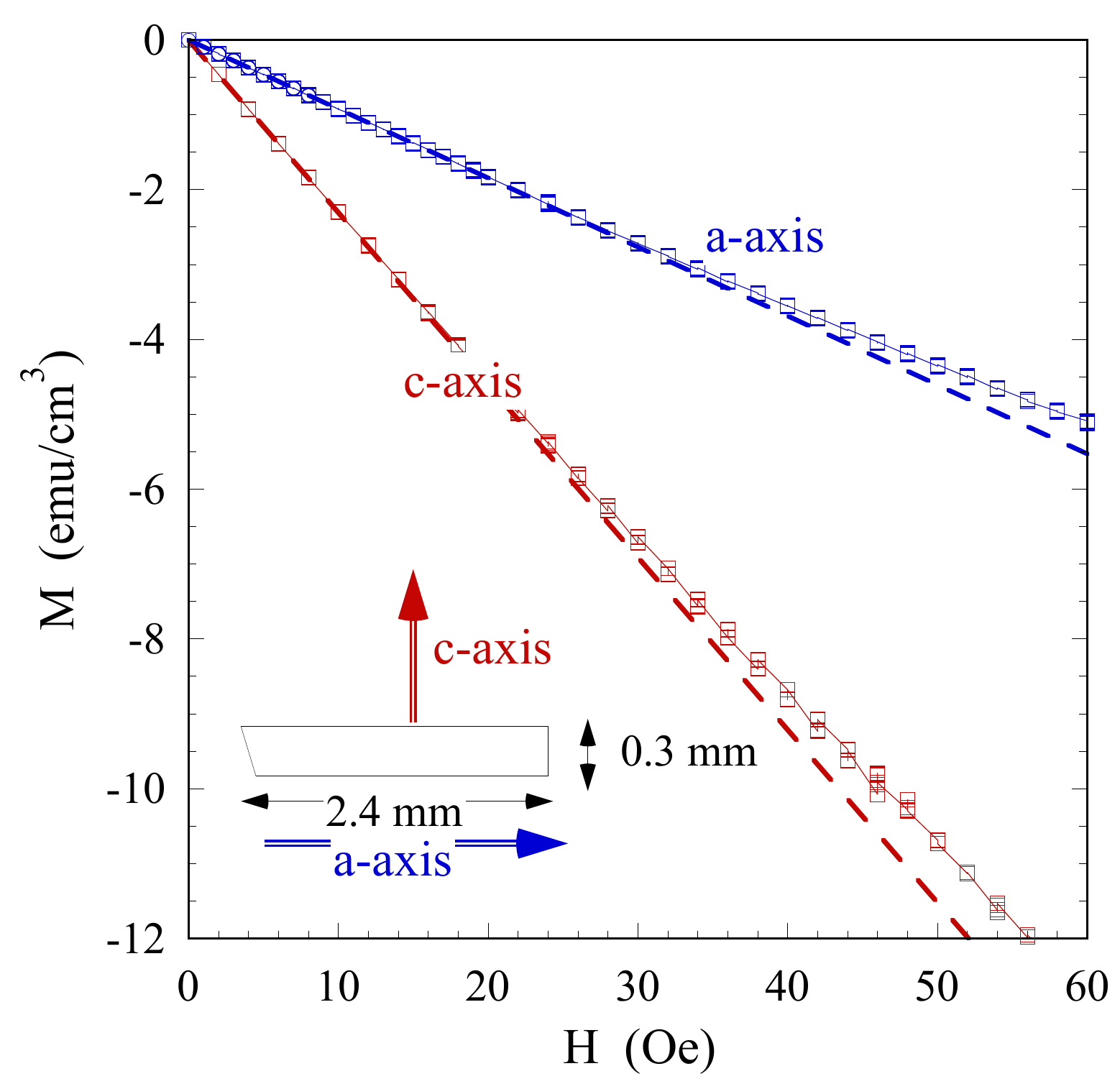} 
\caption{$M$ vs $H$ along the $a$ axis (blue) and $c$ axis (red) for sample 2, taken at 80~mK. The red dashed line is a fit to data for points less than 4 gauss giving an applied susceptibility $M/ H_{\rm applied} =-0.24$ $emu/cm^3$, the blue dashed line for data for points less than 10 gauss giving an applied susceptibility $M/ H_{\rm applied} =-0.09$ $emu/cm^3$. Sample 2 broke while ungluing and the width was reduced to approximately 0.55~mm, the new shape and field directions are shown in the figure.  }
\label{suppl_fig3}
\end{center}
\end{figure}

Notwithstanding, the best strategy for critical low field measurements, as those shown in figure 2 and 4 is simply to plan the experiments at the beginning of a run, and make these measurements first, never increasing the field above 100 Oe.

Figure \ref{suppl_fig1}  shows the initial magnetization curve for sample 1 measured at 80~mK and 1~K along the $a$ axis.
Figures  \ref{suppl_fig2} and \ref{suppl_fig3} show\deleted[id=JP]{s} the initial magnetization curves for sample 2 along the $a$, $b$ and $c$ axis. Both samples had roughly the form of flat platelets as shown in the figures. The mass of sample 1 was 6.88~mg, and the $a$ axis was along the long direction. Sample 2 had a mass of 6.6~mg, and the $b$ axis was along the intermediate direction. Because of the odd shapes, we can only estimate the value of the demagnetization factor using ellipsoids or rectangular shapes. Sample 2 was turned to measure along the $a$ and $c$ axis but unfortunately a portion of the sample broke off. The new mass was 3.9~mg, and the shape of the sample is shown in figure \ref{suppl_fig3}  along with the initial magnetization for the two directions.

From the slope of the initial magnetization curves shown in  \ref{suppl_fig1} and \ref{suppl_fig2}  we find the apparent susceptibility to be - 0.094 and -0.109~emu/cm$^3$  respectively for the two samples. This would imply effective demagnetization coefficients of $N_{\rm effective}$=2 and 3.5 ($N_{\rm effective}$=0.16 and 0.28 in SI units). The values are on the high side of our rough estimates meaning the applied susceptibility is also on the high side, and implies 100 percent shielding. From figure  \ref{suppl_fig3}  the initial slope of (the broken) sample 2 along the $a$ and $c$ axis gives apparent susceptibilities of  - 0.09 and -0.24~$emu/cm^3$ respectively. This implies effective demagnetization coefficients of $N_{\rm effective}$=1.4 and 8.4 ($N_{\rm effective}$=0.11 and 0.67 in SI units). 

We will use the\added[id=JP]{se} effective demagnetization values when correcting the data.

%\newpage

\section{Determination of the thermodynamic critical field}
\begin{figure}[h]
\makeatletter
\renewcommand{\thefigure}{S\@arabic\c@figure}
\makeatother
\begin{center}
\includegraphics[width=0.9\columnwidth]{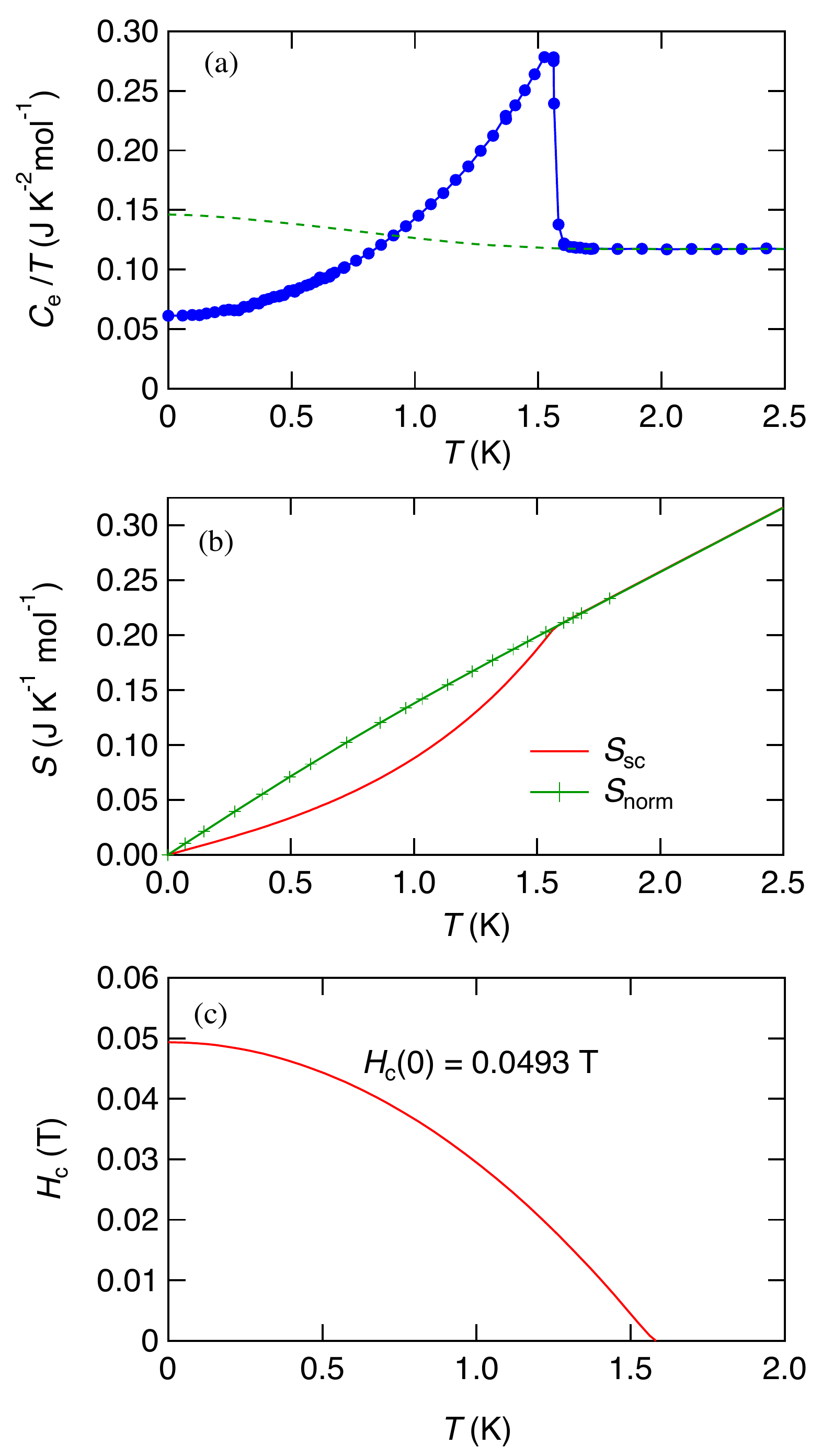}  %{supplemental_figures/specific_heat.pdf}
\caption{(a) Electronic specific heat $C_{\rm el}/T$ of UT$_2$ as function of temperature. $C_{\rm el} = C - C_{\rm ph} - C_{\rm div}$ has been determined following Ref.~\onlinecite{Metz2019} by subtracting the phonon contribution ($C_{\rm ph}\propto T^3$) The dashed line is the expected specific heat in the normal state taking into account the entropy balance. (b) Temperature dependence of the entropy in the normal (extrapolated below $T_c$) and superconducting state. (c) Temperature dependence of the thermodynamic critical field $H_{\rm c}$ deduced from the difference in the free energy  $\Delta F = F_{\rm norm} - F_{\rm sc} = \int_{T}^{T_c} \Delta S(T) dT = H_{\rm c}^2 / 2\mu_0$.   }
\label{suppl_fig4}
\end{center}
\end{figure}

The specific heat has been measured with a relaxation method down to 100~mK in zero magnetic field on a different sample.  The electronic specific heat, shown in Fig.~\ref{suppl_fig4}a) $C_{\rm el} = C - C_{\rm ph}  - C_{\rm div}$   has been determined in the same way than in Ref.~\onlinecite{Metz2019} by subtracting a phonon contribution ($C_{\rm ph} \propto T^3$) and a small to low temperatures diverging term  $C_{\rm div} \propto T^{1-\alpha}$. The dashed line is the expected normal state specific heat taking into account the entropy balance. Figure~\added[id=JP]{\ref{suppl_fig4}}b) shows the temperature dependence of the entropy in the normal and superconducting state. The thermodynamic critical field $H_c$ can be extracted from the difference in the free energy  $\Delta F = F_{\rm norm} - F_{\rm sc} = \int_{T}^{T_c} \Delta S(T) dT = H_{\rm c}^2 / 2\mu_0$ of the normal state and the superconducting state\added[id=JP]{s}.  From this analysis we find the isotropic thermodynamic critical field $H_c (0)= 0.0493$~T.

%\clearpage
\section{Hysteresis loops close to the superconducting transition}

\begin{figure}[h]
\makeatletter
\renewcommand{\thefigure}{S\@arabic\c@figure}
\makeatother
\begin{center}
\includegraphics[width=0.9\columnwidth]{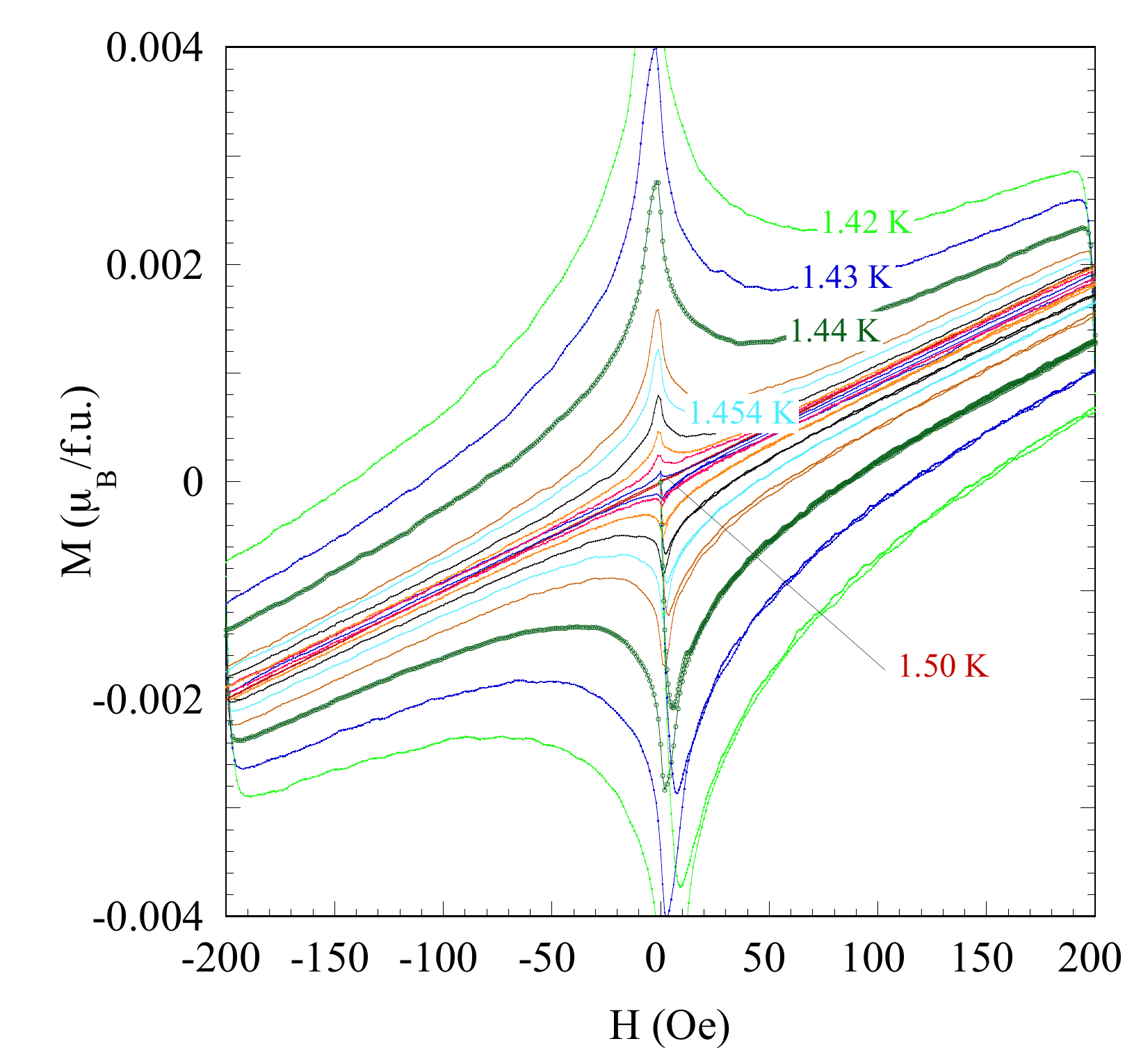} %{supplemental_figures/suppl_fig1.png}
\caption{Series of minor hysteresis loops for UTe$_2$ with the field direction along the $a$ axis at temperatures close to $T_{\rm c}$. The hysteresis loops have a pronounced diamond shape indicating strong pinning. As the temperature approaches  $T_{\rm c}$, the hysteresis loops collapse onto the relatively large paramagnetic background\deleted[id=JP]{response}. No ferromagnetic behavior is observed.}
\label{suppl_fig5}
\end{center}
\end{figure}

Figure \ref{suppl_fig5} displays several minor hysteresis loops very close to the superconducting transition $T_c$ for sample 1 with the field applied along the $a$ axis. Clearly, in the superconducting state below $T_c$ the magnetization shows pronounced diamond shape indicating \deleted[id=JP]{the} strong pinning on top of a \replaced[id=JP]{large}{strong} paramagnetic background. Above $T_c$ only the paramagnetic signal is observed. No indication of any ferromagnetic signal is \replaced[id=JP]{detected}{observed}. 

\begin{figure}[t]
\makeatletter
\renewcommand{\thefigure}{S\@arabic\c@figure}
\makeatother
\begin{center}
\includegraphics[width=0.9\columnwidth]{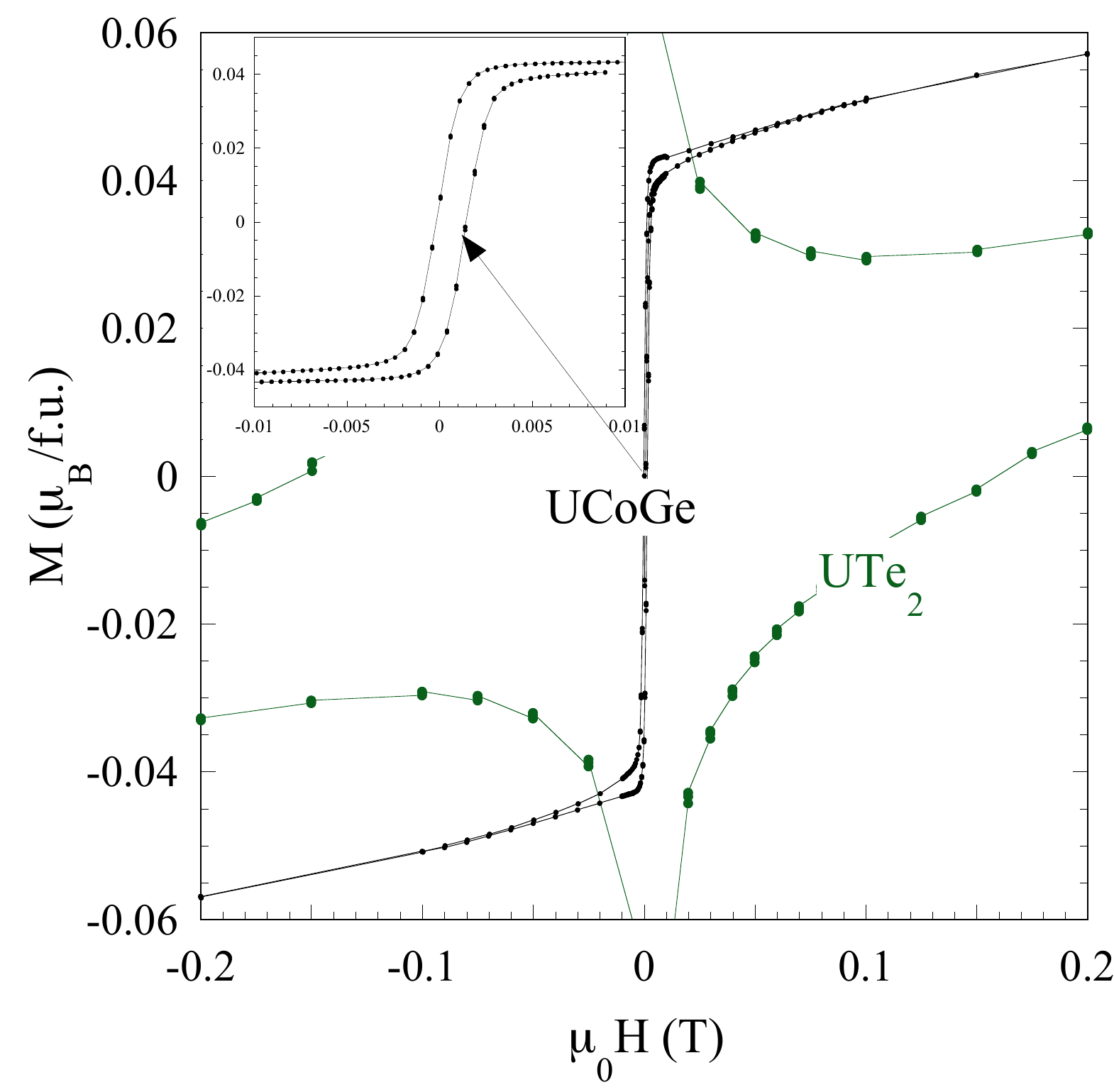}  %{supplemental_figures/suppl_fig2.png}
\caption{Hysteresis loops for UTe$_2$ at 1~K compared with UCoGe  at $T=70$~mK. The pure superconducting hysteresis response of UTe$_2$ is in contrast to UCoGe where the superconducting and ferromagnetic signals are fused together with the FM response dominating \cite{Paulsen2012}. In fact, at first glance UCoGe does not look superconducting at all. However, the diamond shape response for UCoGe can be revealed by subtracting the hysteresis measured just above the superconducting transition from the low temperature data \cite{Deguchi2010}. }
\label{suppl_fig6}
\end{center}
\end{figure}

\begin{figure}[t]
\makeatletter
\renewcommand{\thefigure}{S\@arabic\c@figure}
\makeatother
\begin{center}
\includegraphics[width=0.9\columnwidth]{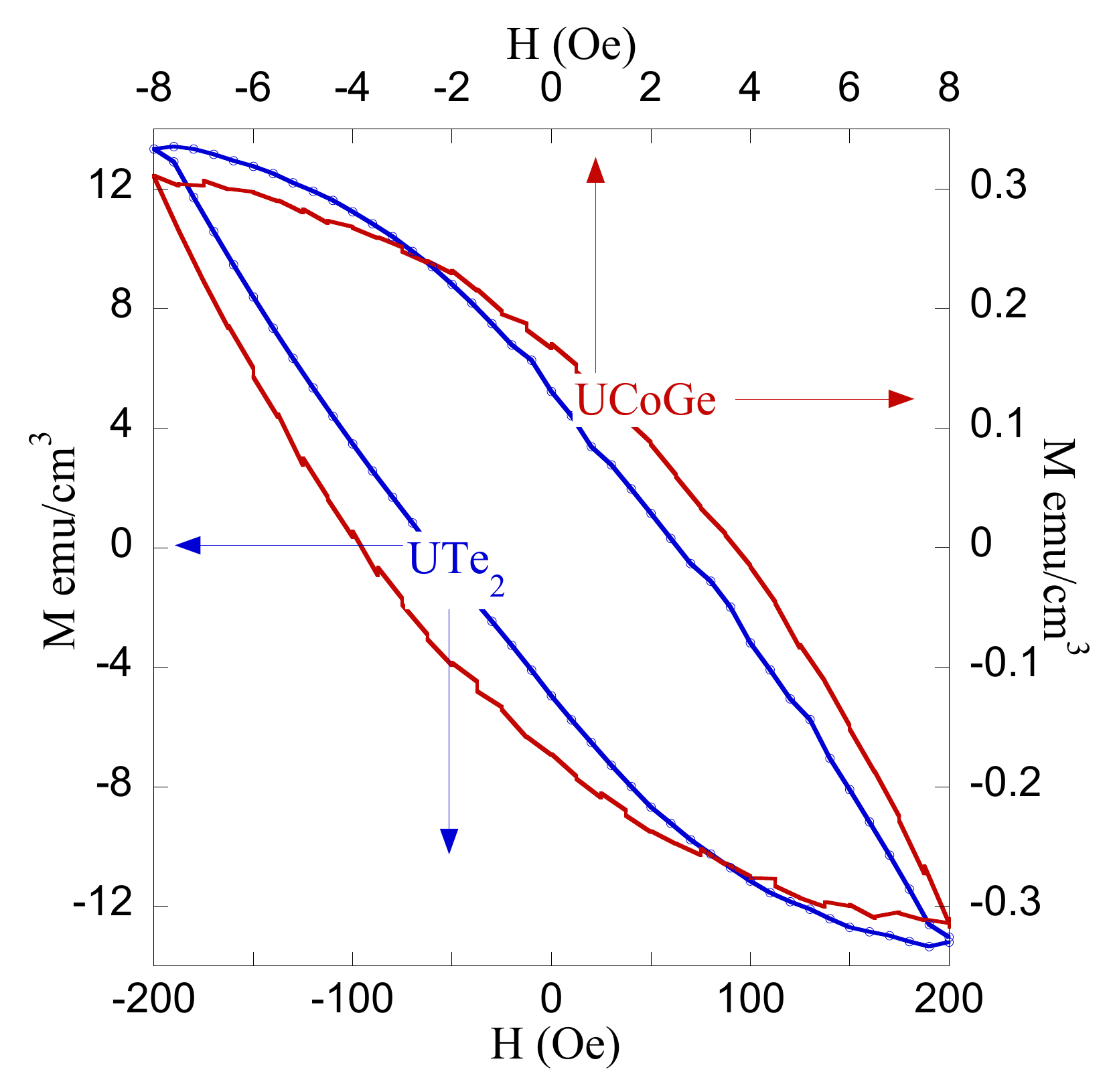} % {supplemental_figures/suppl_fig3.png}
\caption{Minor hysteresis loops for UTe$_2$ measured at 800~mK, and UCoGe measured at 70~mK. Another way to see the pure superconducting response for UCoGe is to measure minor hysteresis loops using fields smaller than the FM coercive field as shown here. Also shown is a minor loop for UTe$_2$ but on vastly different scale. The form of these curves can be nicely fitted using the Bean critical state model \cite{Bean1964}.}
\label{suppl_fig7}
\end{center}
\end{figure}

\highlight[id=JP, comment = {Carley, there is a label (c) on this figure, useless, which should be removed !}]{In Fig.~\ref{suppl_fig6} } we compare hysteresis loops for UTe$_2$ at 1~K   with that of UCoGe measured at 70~mK. The shape of the hysteresis loops for UCoGe, while in the coexisting ferromagnetic and superconducting state, appears to be dominated by the ferromagnetic response \cite{Paulsen2012}. As shown in Refs.~\onlinecite{Deguchi2010, Paulsen2012} the characteristic diamond shape of the magnetization of a type-II superconductor can be obtained after the subtraction of the ferromagnetic contribution above the superconducting transition from the low temperature data. This gives clear evidence for the coexistence of superconductivity and ferromagnetism in UCoGe while in UTe$_2$ no ferromagnetism occurs down to the lowest temperatures.  
The pure superconducting response for both compounds can be seen in Fig.~\ref{suppl_fig7} which shows minor hysteresis loops. This time, for UCoGe, the applied magnetic field was kept smaller than the ferromagnetic coercive field of the sample ($\sim 10$~Oe).

%\clearpage

\section{Determination of lower critical field $H_{\rm c1}$}

When type-II superconductors have strong pinning, flux enters the sample almost asymptotically, as seen in figures \replaced[id=JP]{ \ref{suppl_fig1}, \ref{suppl_fig2} and \ref{suppl_fig3}}{S1, S2 and S3}. This makes determination of $H_{\rm c1}$ difficult. In addition, because the samples used in this study were not perfect ellipsoids, the field \replaced[id=JP]{enters}{will enter} the sample sooner around the edges due to demagnetization effects, further obscuring the real $H_{\rm c1}$. 
To overcome this \added[id=JP]{problem}, different methods were tried in order to \replaced[id=JP]{best determine}{measure} the lower critical field $H_{\rm c1}$.

\begin{figure}[h]
\makeatletter
\renewcommand{\thefigure}{S\@arabic\c@figure}
\makeatother
\begin{center}
\includegraphics[width=0.9\columnwidth]{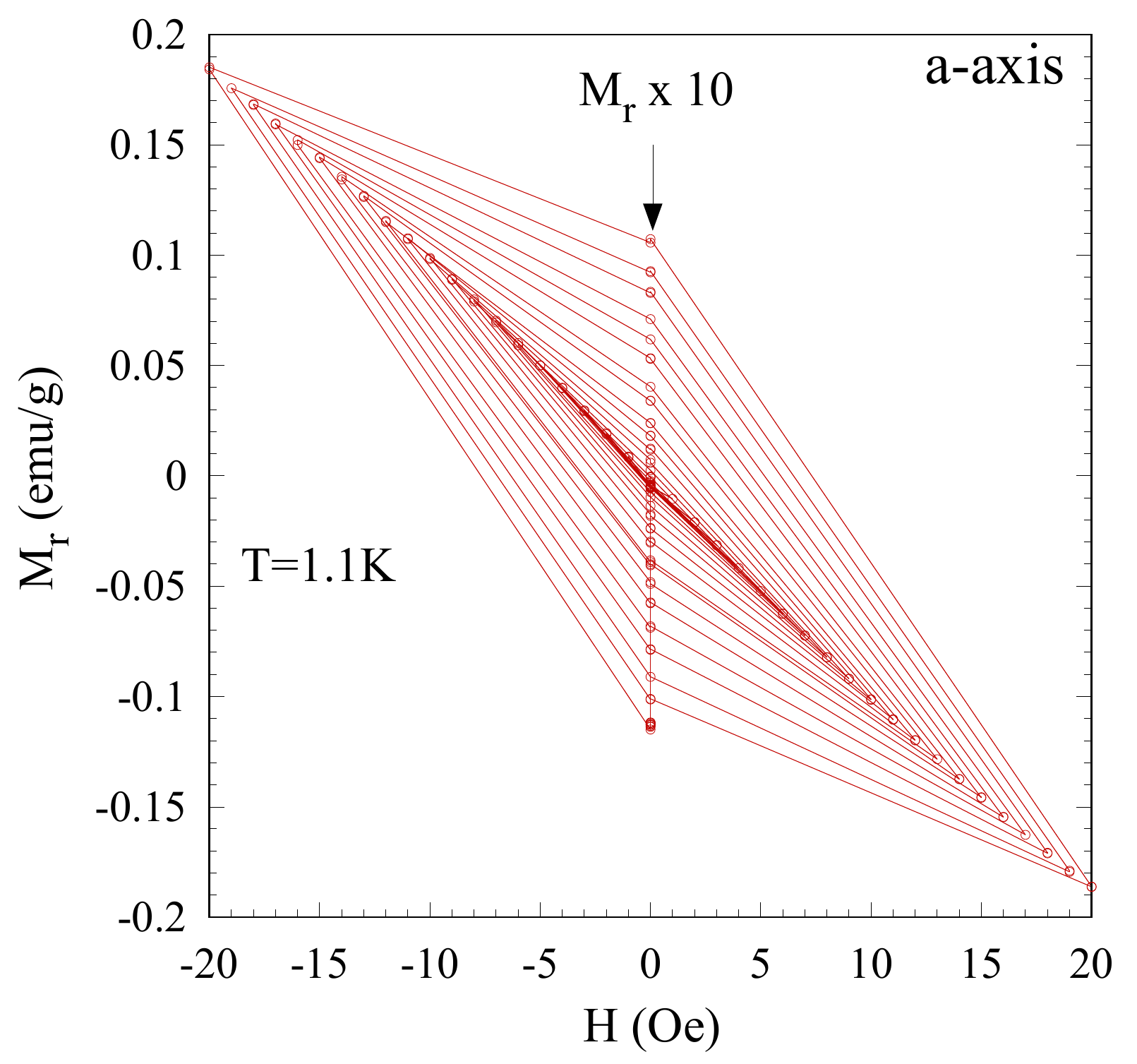} 
\caption{An example of a series of increasing minor hysteresis loops taken at 1.1~K used for the determination of  $H_{\rm c1}$ along the $a$ axis. This data was taken by first zero field cooling the sample to 1.1~K.  Then a field of 1 Oe  was applied, the magnetization was measured, the field was removed, the magnetization was measured, the field was reversed to -1 Oe,  measured, then returned to zero and $M\added[id=JP]{_{\rm r}}$ was measured. The field was then increased to 2~Oe, then back to zero, then to -2 and back to zero and so on, measuring $M\added[id=JP]{_{\rm r}}$ at each field change, and systematically increasing the hysteresis loops by 1~Oe steps up to 20~Oe. While $H< H_{\rm c1}$ the cycles are
reversible, and there is no remanent magnetization detected when the field is reduced to zero. However, when $H_{\rm c1}$ is exceeded, then flux begins to enter the sample and the magnetization deviates from the 100\% shielding. When the field is returned to zero, one half of the flux is trapped in
the sample, and a remanent magnetization appears (in the plot it has been multiplied by 10). The procedure was repeated for different constant temperatures.}
\label{suppl_fig8}
\end{center}
\end{figure}

The method used for the data shown in Fig.~\replaced[id=JP]{4d}{4b} in the main text is based on measurements of the remanent magnetization \added[id=JP]{$M_{\rm r}$}, and is outlined below and demonstrated in the plots. Using the remanent magnetization when searching for $H_{\rm c1}$ with bulk measurements has advantages. Normally when using the magnetization, one subtracts off the linear part, and the difference is used to observe $H_{\rm c1}$. At high fields this means that we are \replaced[id=JP]{subtracting two large numbers in order to extract}{subtracting a large number from a large number to get} a small difference. This is not necessary when using the remanent magnetization. In addition the remanent \added[id=JP]{magnetization} is measured in zero field,  \replaced[id=JP]{hence with }{thus} less noise,  \replaced[id=JP]{so that }{and} higher \replaced[id=JP]{SQUID}{squid} gains can be used. Nevertheless we also show here that using $M$ vs $H$ \replaced[id=JP]{can give}{gives} the same result.

\begin{figure}[h]
\makeatletter
\renewcommand{\thefigure}{S\@arabic\c@figure}
\makeatother
\begin{center}
\includegraphics[width=0.9\columnwidth]{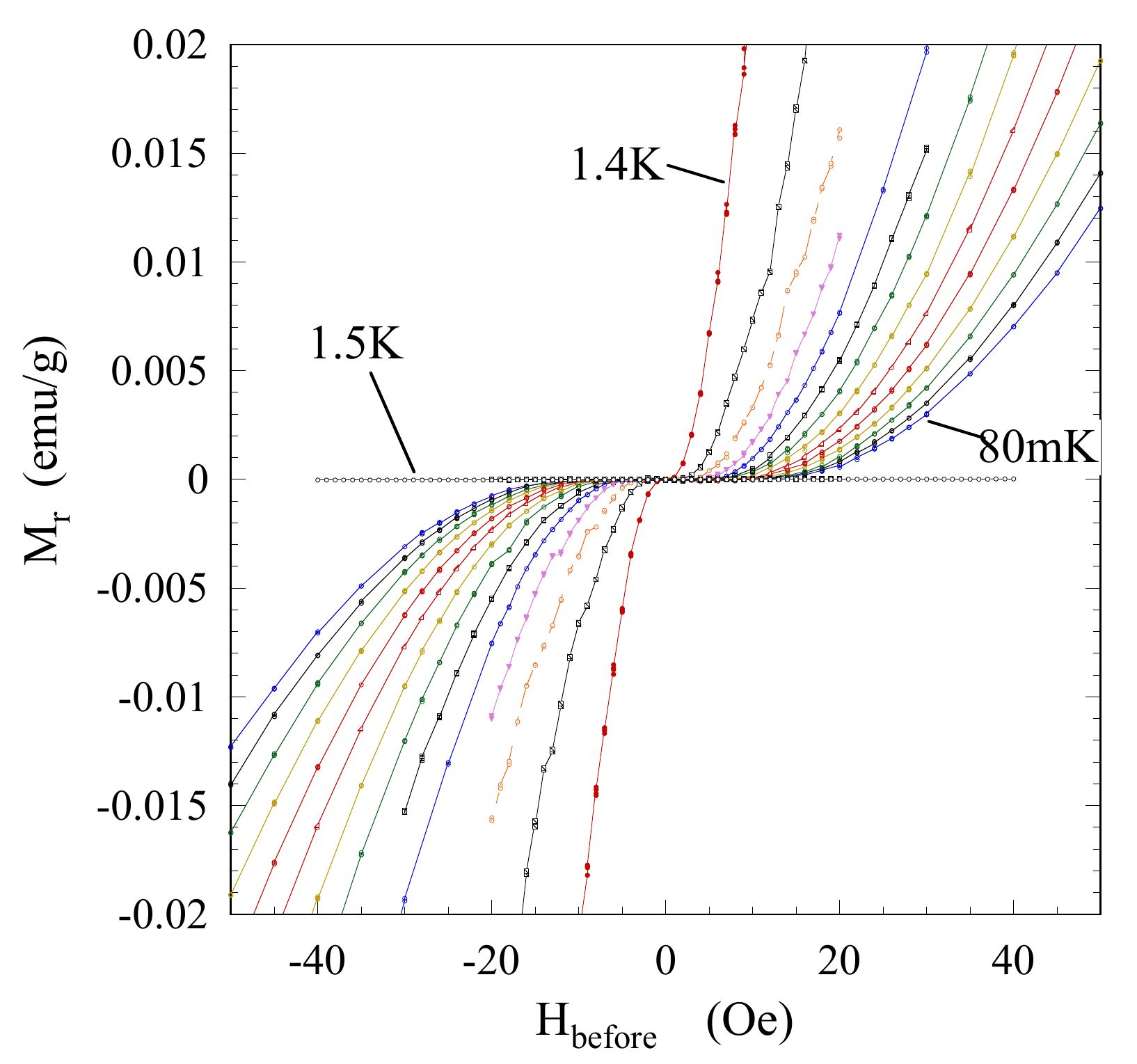} 
\caption{ \replaced[id=JP]{R}{A plot the r}emanent magnetization along the $a$ axis measured at various temperatures vs $H_{\rm before}$. The remanent  magnetization was measured in zero field as shown in \ref{suppl_fig8}, but it is plotted in the figure as a function of $H_{\rm before}$, i.e. the last field value before the field was reduced to zero and the remanence was measured. }
\label{suppl_fig9}
\end{center}
\end{figure}

An example of the method used to measure the remanent magnetization is shown if \highlight[id=JP, comment={Carley, on figures S8, S9, S11, S12, S14, S15, I believe we should replace "remanent M " by $M_r$, in order to keep consistent notations everywhere.}]{figure \ref{suppl_fig8} for sample 1} along the $a$ axis at 1.1~K. The plot shows a sequence of ever increasing hysteresis loops, in steps of 1~Oe, and the resulting remanent magnetization when the field is reduce\added[id=JP]{d} to zero. Figure \ref{suppl_fig9} shows the results of measurements of the remanent magnetization made at various constant temperatures, plotted in the figure as a function of $H_{\rm before}$, i.e. the last field value before the field was reduced to zero and the remanence was measured.

According to the Bean critical state model  (here for a slab\added[id=JP]{-}shaped sample), when flux enters a type-II superconductor, the deviation from perfect screening is $\propto H^2$: $-4\pi M=H-H^2/2H^*$ and it retains this flux when the field is removed, giving rise to a remanent magnetization $4\pi M_{\rm r}=H\added[id=JP]{_{\rm before}}^2/4H^*$, w\added[id=JP]{h}ere $H^*= \pi J_cD/5$  using practical units\added[id=JP]{:} $H$ in Oe, $D$ the thickness of the slab in cm, and  $J_c$ the critical current  i.e the current that impedes the field from penetrating into the sample\added[id=JP]{, in A/cm$^2$}.

%\newpage

In order to make use of these ideas to estimate $H_{\rm c1}$, we assume two regimes above and below  $H_{\rm c1}$:  

\begin{itemize}
\item  if $H < H_{\rm c1}$: 
\begin{eqnarray*}
 4\pi M&=&-H \\
  4\pi M_{\rm r}&=&0
 \end{eqnarray*}
\item  and when $H_{\rm c1} < H < H^*$:
\begin{eqnarray*}
4\pi M&=&-H  + (H-H_{\rm c1})^2/2H^* \\				
4\pi M_{\rm r}&=&(H\added[id=JP]{_{\rm before}}-H_{\rm c1})^2/4H^*	
 \end{eqnarray*}
\end{itemize}

\begin{figure}[b]
\makeatletter
\renewcommand{\thefigure}{S\@arabic\c@figure}
\makeatother
\begin{center}
\includegraphics[width=0.7\columnwidth]{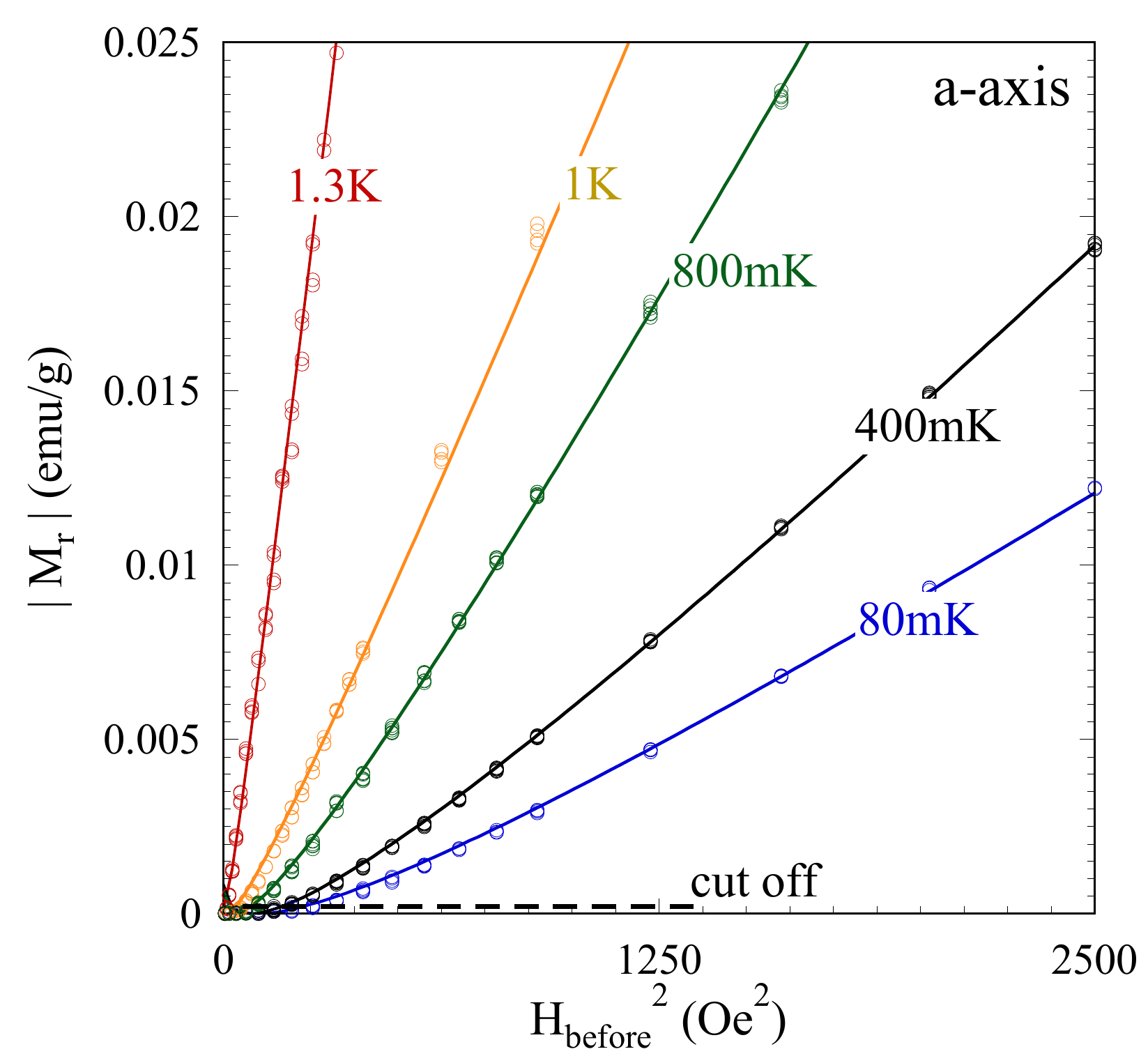} 
\caption{A few of the curves from figure \protect{\ref{suppl_fig9}} plotted as $M\added[id=JP]{_{\rm r}}$ vs  
$H\added[id=JP]{_{\rm {before}}}^2$, the square of the last field value before the field was reduced to zero. The solid lines are fits to the data using the form $(H\added[id=JP]{_{\rm before}}-H_{\rm c1})^2 $. The data below the dashed cut off line were not used in the fits.
}
\label{suppl_fig10}
\end{center}
\end{figure}

\begin{figure}[h]
\makeatletter
\renewcommand{\thefigure}{S\@arabic\c@figure}
\makeatother
\begin{center}
\includegraphics[width=0.75\columnwidth]{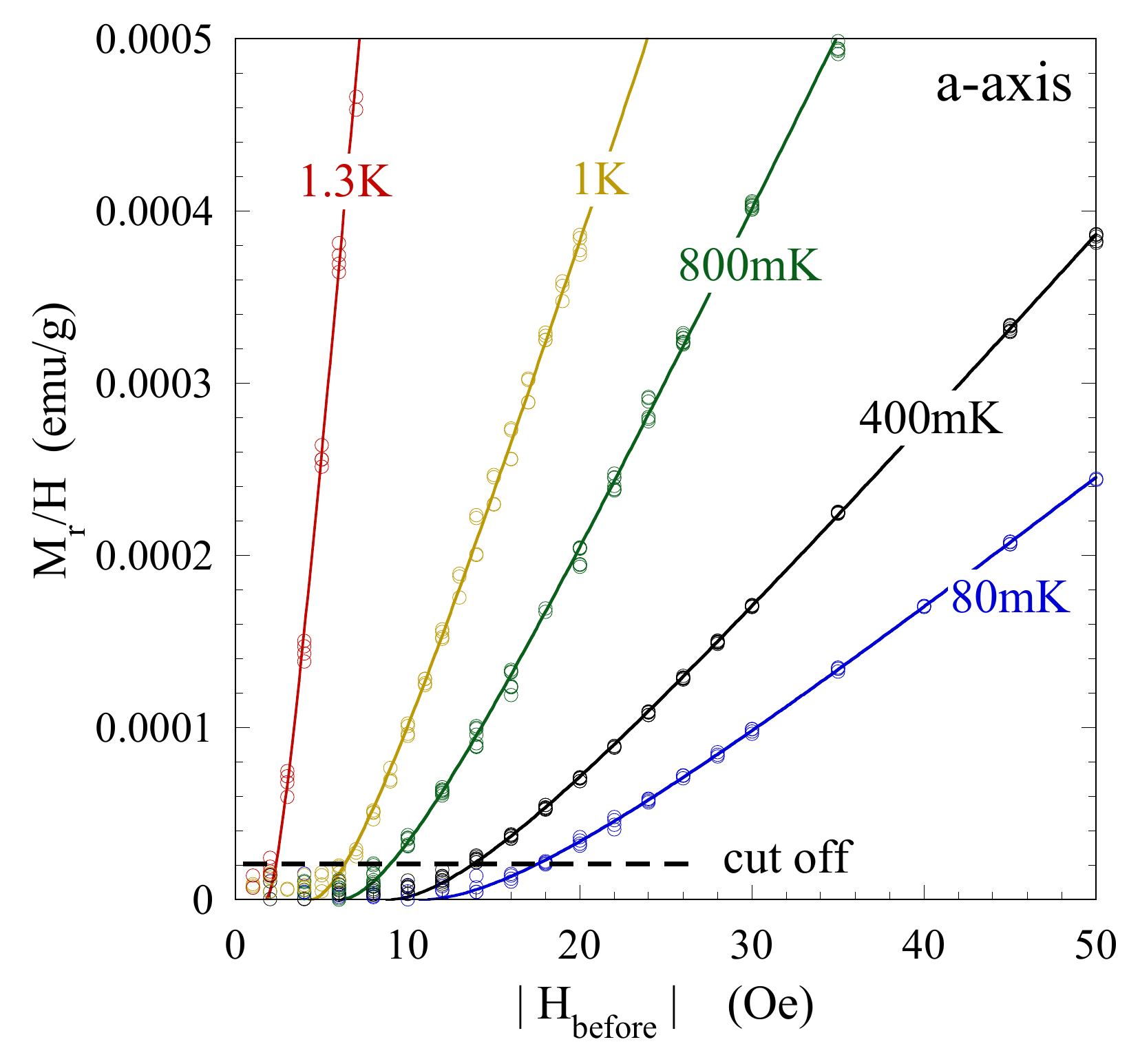} 
\caption{ Example of the same data plotted as remanent $M\added[id=JP]{_{\rm r}}/H\added[id=JP]{_ {\rm before}}$ vs the absolute value of $H_{\rm before}$. The solid lines are fits to the data using the form $(H\added[id=JP]{_{\rm before}}-H_{\rm c1})^2/H\added[id=JP]{^{\rm *}} $. The data below the dashed cut off line were not used in the fits.
}
\label{suppl_fig11}
\end{center}
\end{figure}

\begin{figure}[b]
\makeatletter
\renewcommand{\thefigure}{S\@arabic\c@figure}
\makeatother
\begin{center}
\includegraphics[width=0.75\columnwidth]{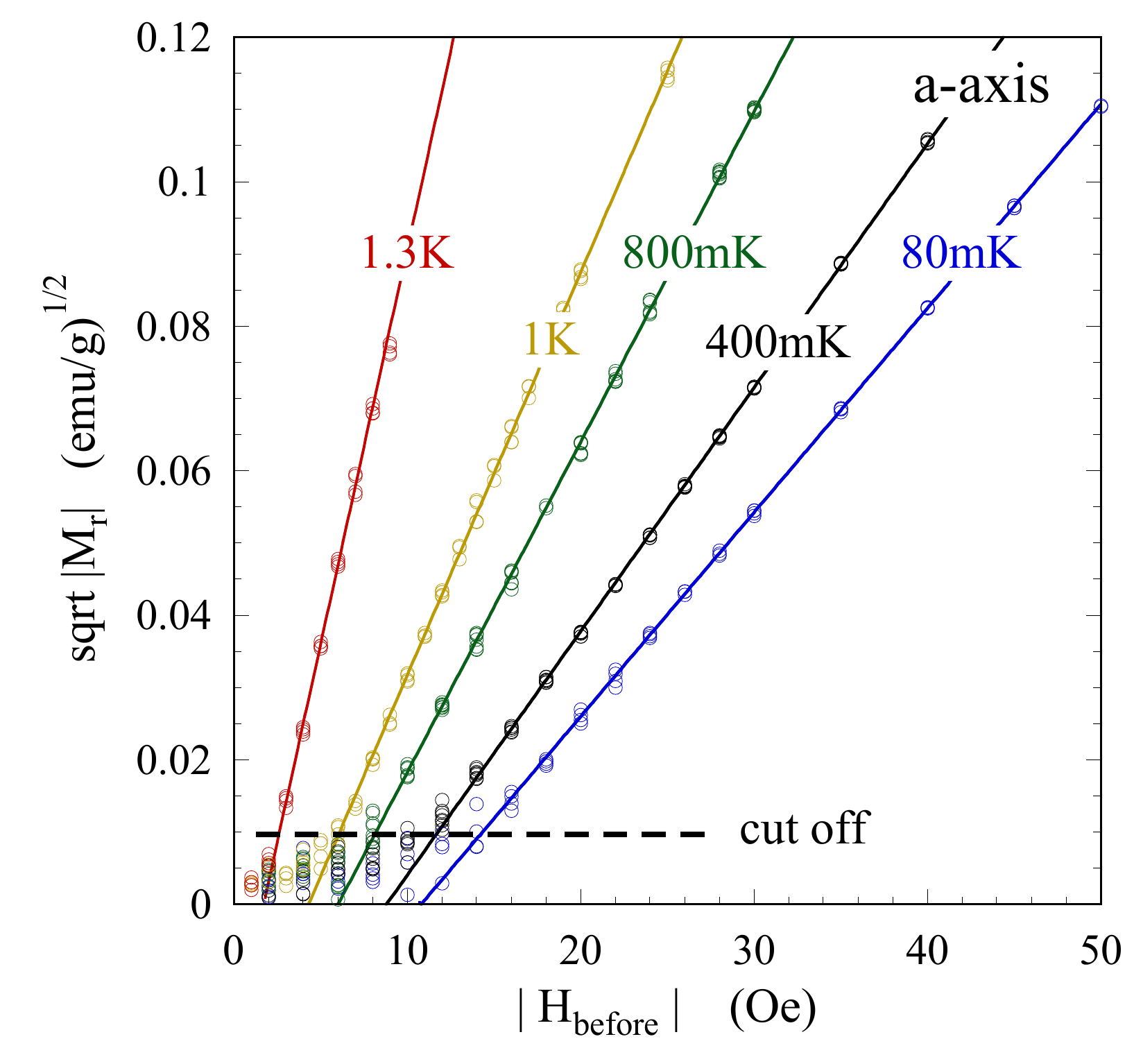} 
\caption{ The data plotted this time as $\sqrt{|M\added[id=JP]{_{\rm r}}|}$  vs absolute value of $H_{\rm before}$. The solid lines are linear fits to the data above the dashed cut off line. From this figure $H_{\rm c1}$ can be seen as the intersection of the linear fits with the $x$ axis.
}
\label{suppl_fig12}
\end{center}
\end{figure}

We focus on the last equation for $M_{\rm r}$, and ignoring demagnetization corrections for the moment consider three ways to exploit this relation to analyze the data shown in Fig.~\ref{suppl_fig9}: 

1) We can plot the data vs $H\added[id=JP]{_{\rm before}}^2$ as shown in  \highlight[id = JP, comment = {Carley, on the figure, shouldn't we use $H_{\rm before}$, not $H_{\rm b}$}]{Fig.~\ref{suppl_fig10}}, and then fit it to $(H\added[id=JP]{_{\rm before}}-H_{\rm c1})^2$. 

2) We can plot the data as $M_{\rm r}/H\added[id=JP]{_{\rm before}}$ vs $ H\added[id=JP]{_{\rm before}}$, and fit the data to $(H\added[id=JP]{_{\rm before}}-H_{\rm c1})^2/H\added[id=JP]{_{\rm before}}$ as shown in figure \ref{suppl_fig11}. 

3) We plot the data as $\sqrt{M_{\rm r}}$ vs $H\added[id=JP]{_{\rm before}}$, and simply fit the data using a linear fit as shown in Fig.~\ref{suppl_fig12}. 

\begin{figure}[t]
\makeatletter
\renewcommand{\thefigure}{S\@arabic\c@figure}
\makeatother
\begin{center}
\includegraphics[width=0.8\columnwidth]{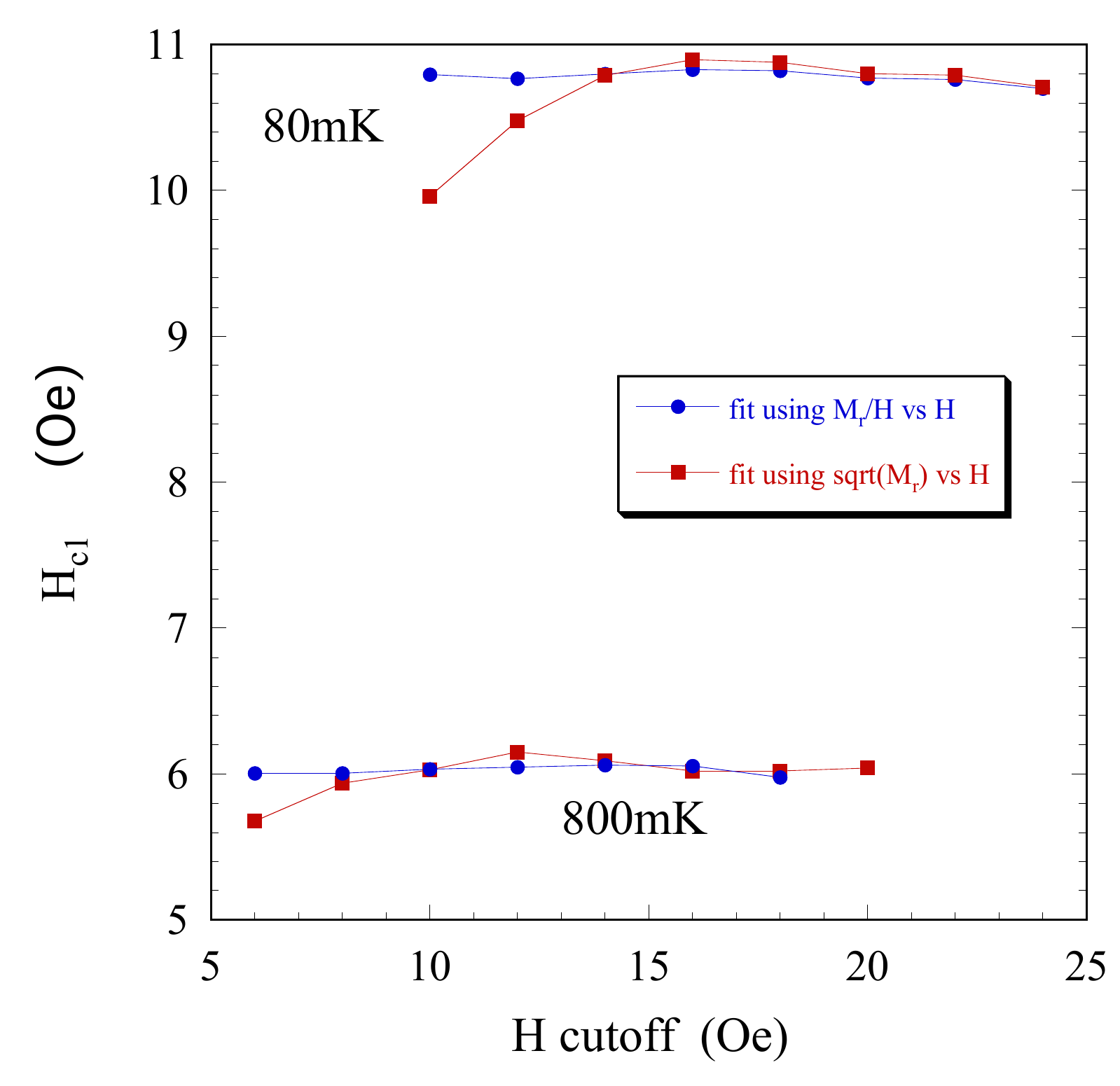} 
\caption{ Example of the sensitivity of fits for $H_{\rm c1}$ vs the cutoff field used for the fits, at 80~mK and 800~mK for the $a$ axis. }
\label{suppl_fig13}
\end{center}
\end{figure}

The three fitting procedures give more or less the same result, however fitting the data to  $(H\added[id=JP]{_{\rm before}}-H_{\rm c1})/H\added[id=JP]{_{\rm before}} $ \highlight[id=JP, comment = {Carley, on figure \ref{suppl_fig13}, it seems that it is the fit of $\sqrt{M}$ vs $H_{\rm before}$ which deviates more strongly at low fields ? Or is there some inversion in the legend on the figure ?}] is more robust at lower fields as shown in figure \ref{suppl_fig13}.

\begin{figure}[h]
\makeatletter
\renewcommand{\thefigure}{S\@arabic\c@figure}
\makeatother
\begin{center}
\includegraphics[width=0.75\columnwidth]{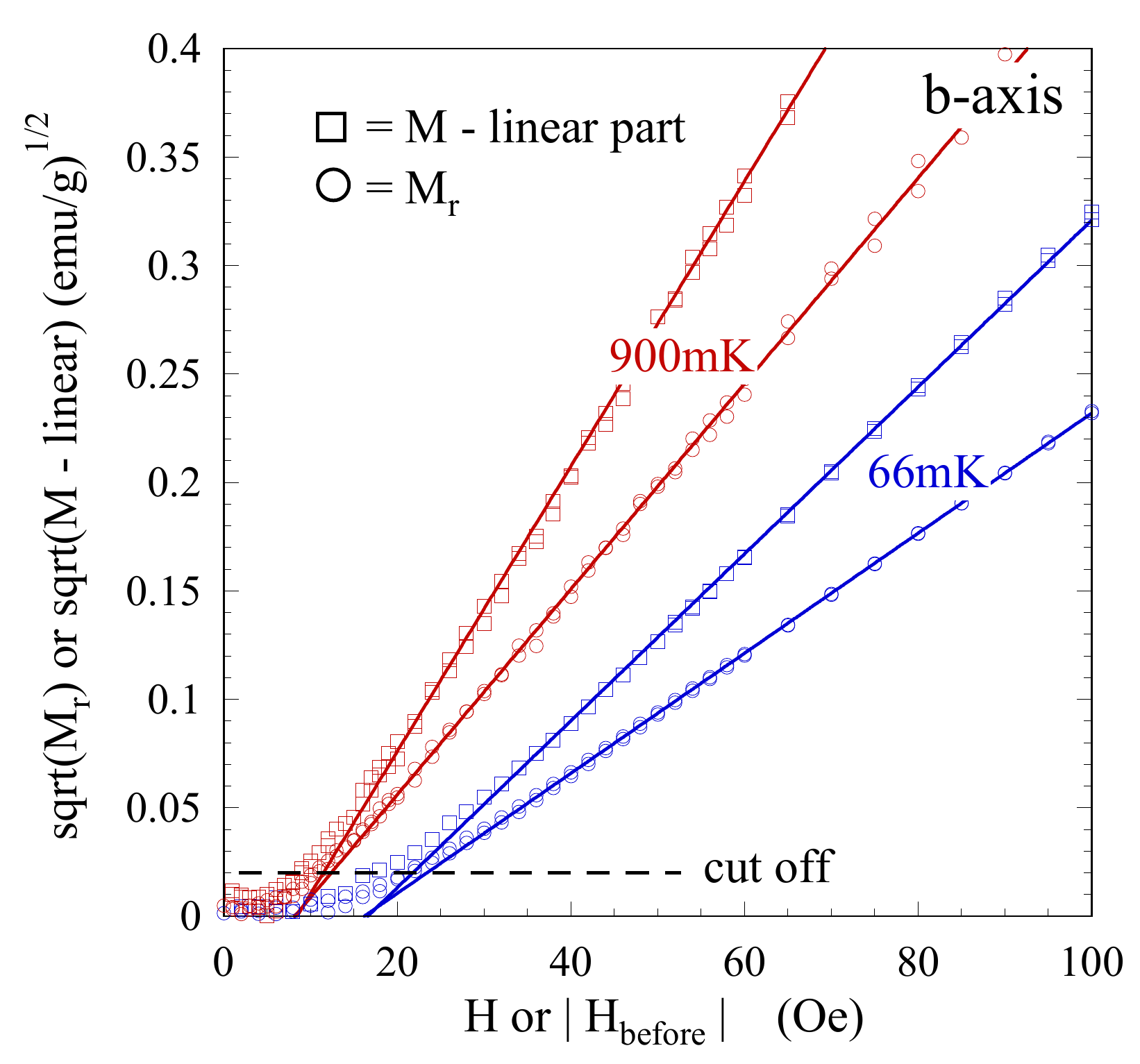} 
\caption{ Plot of the square root of the magnetization minus the linear fit to the data at low field vs $H$ compared to the $\sqrt{ |M\added[id=JP]{_{\rm r}}|}$ vs  $H_{\rm before}$. This is data from the second sample measured along the $b$ axis. The solid lines are linear fits to the data above the cut off. The figure shows that  $H_{\rm c1}$ determined from the magnetization $M$ or the $M\added[id=JP]{_{\rm r}}$ are equivalent.}
\label{suppl_fig14}
\end{center}
\end{figure}

Figure \ref{suppl_fig14}  is an example of data from the second sample measured along the $b$ axis at 66 and 900~mK. The square symbols are the magnetization data minus the linear fit to the data at low field, (i.e. we subtract the dashed line shown in Fig.~\ref{suppl_fig2} ). The circles are the remanent data. The data are plotted as sqrt(M-linear part) and sqrt( $|M_{\rm r}|)$ vs  $H_{\rm before}$. The solid lines are linear fits to the data above the cut off. The figure shows that  $H_{\rm c1}$ determined from the magnetization $M$ and from the remanent magnetization $M_{\rm r}$ are equivalent within $\pm 1$~Oe.

\begin{figure}[h]
\makeatletter
\renewcommand{\thefigure}{S\@arabic\c@figure}
\makeatother
\begin{center}
\includegraphics[width=0.75\columnwidth]{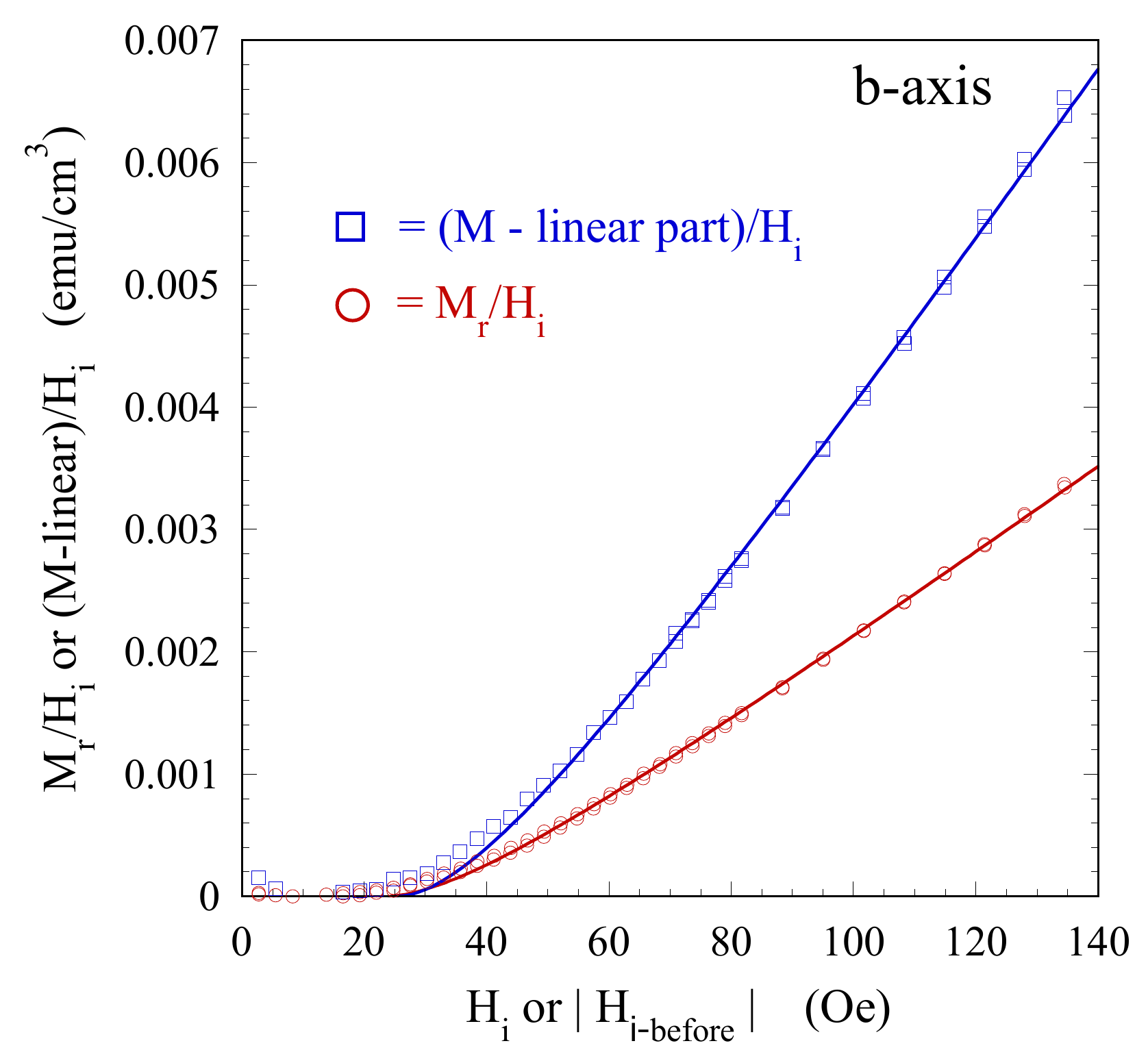} 
\caption{ The same data for the second sample along the $b$ axis at 66~mK as shown in figure \ref{suppl_fig14}, but here plotted as $M\added[id=JP]{_{\rm r}}/H_{\rm i}$ or $(M-linear)/H_{\rm i}$,  vs $H_{\rm i}$, where $H_{\rm i}$ is the internal field after correcting for demagnetization effects.  From the figure it can be seen that the remanent slope is approximately 1/2 the slope found from the magnetization, as it should in accordance to the critical state model.}
\label{suppl_fig15}
\end{center}
\end{figure}

An example with demagnetization corrections is shown in Figure \ref{suppl_fig15}  for data from the second sample measured along the $b$ axis at 66~mK. The data have been corrected for demagnetization effects using the $N_{\rm effective}$= 3.5 determined from figure  \ref{suppl_fig2}. The square symbols are the magnetization data minus the linear fit to the data at low field and the circles are the remanent data. The data are plotted as ($M$-linear part)/$H_{\rm i}$ and $|M_{\rm r}|/H_{\rm i-before}$ vs  $H_{\rm i}$ or $H_{\rm i-before}$. Correcting for demagnetization effects gives a critical field about 1.4 times larger along the $b$ axis and about 1.2 times larger along the $a$ axis. The corrected values are used in Fig.~4 of the main text.

\begin{figure}[b]
\makeatletter
\renewcommand{\thefigure}{S\@arabic\c@figure}
\makeatother
\begin{center}
\includegraphics[width=0.75\columnwidth]{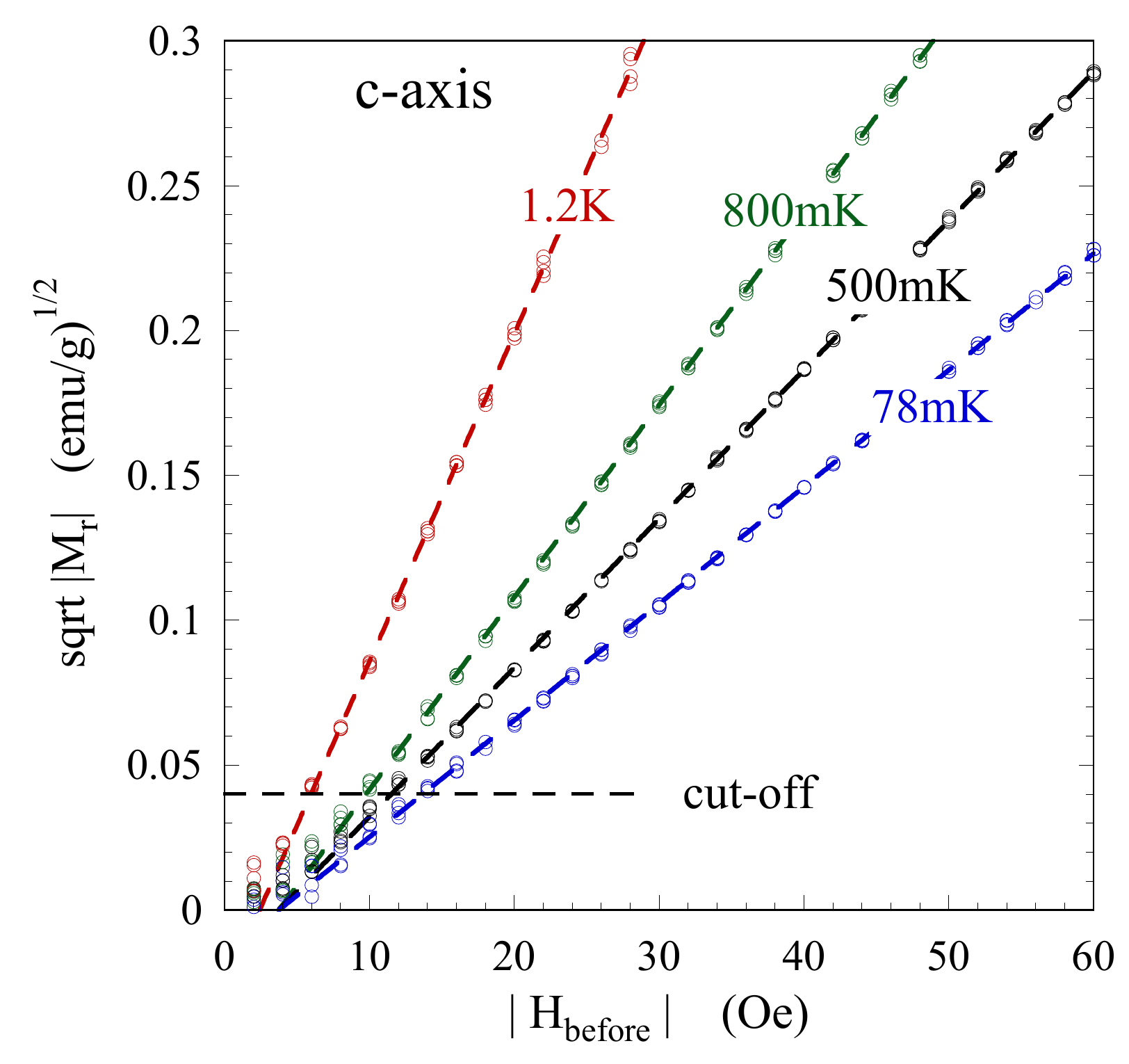} 
\caption{ The remanent magnetization for sample 2 measured along $c$ axis plotted as sqrt$( |M_{\rm r}|)$ vs $H_{\rm before}$. The dashed lines are fits to the data above the cut-off. Even for measurements perpendicular to the flattest dimension, above the cut-off the field still penetrates to a very good approximation as $(H-H_{\rm c1})^2$. }
\label{suppl_fig16}
\end{center}
\end{figure}

\begin{figure}[h]
\makeatletter
\renewcommand{\thefigure}{S\@arabic\c@figure}
\makeatother
\begin{center}
\includegraphics[width=0.75\columnwidth]{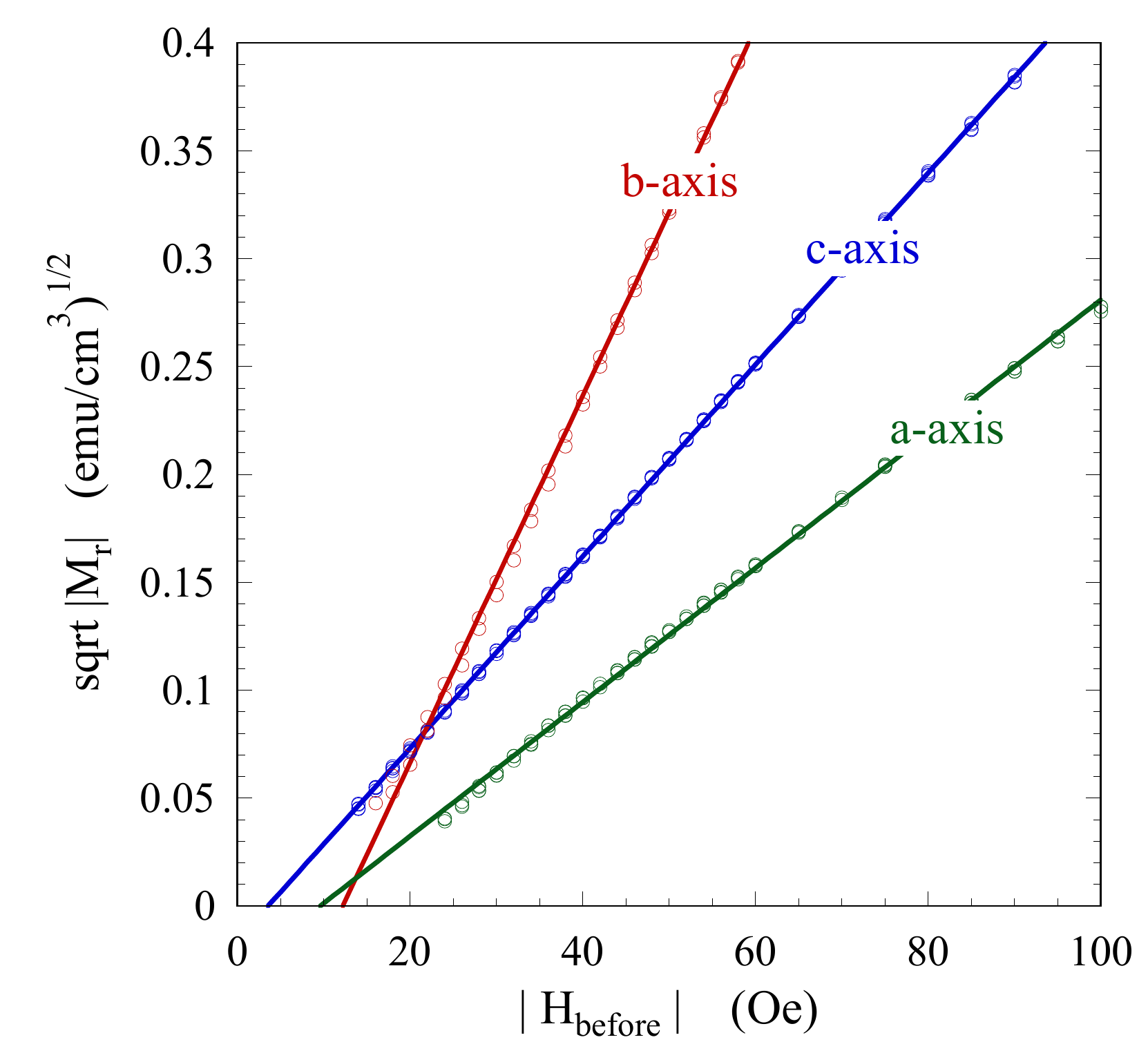} 
\caption{ Sample 2 measured along the $a$,  $b$ and $c$ axis at 300~mK plotted as $\sqrt{ |M_{\rm r}|}$ vs $H_{\rm before}$ (the low field points have been cut  for clarity). Without \added[id=JP]{demagnetization} corrections,  \added[id=JP]{a much smaller value of }$H_{\rm c1}$  \replaced[id=JP]{along}{for} the $c$ axis  \added[id=JP]{would have been obtained.}}
\label{suppl_fig17}
\end{center}
\end{figure}

\begin{figure}[h]
\makeatletter
\renewcommand{\thefigure}{S\@arabic\c@figure}
\makeatother
\begin{center}
\includegraphics[width=0.75\columnwidth]{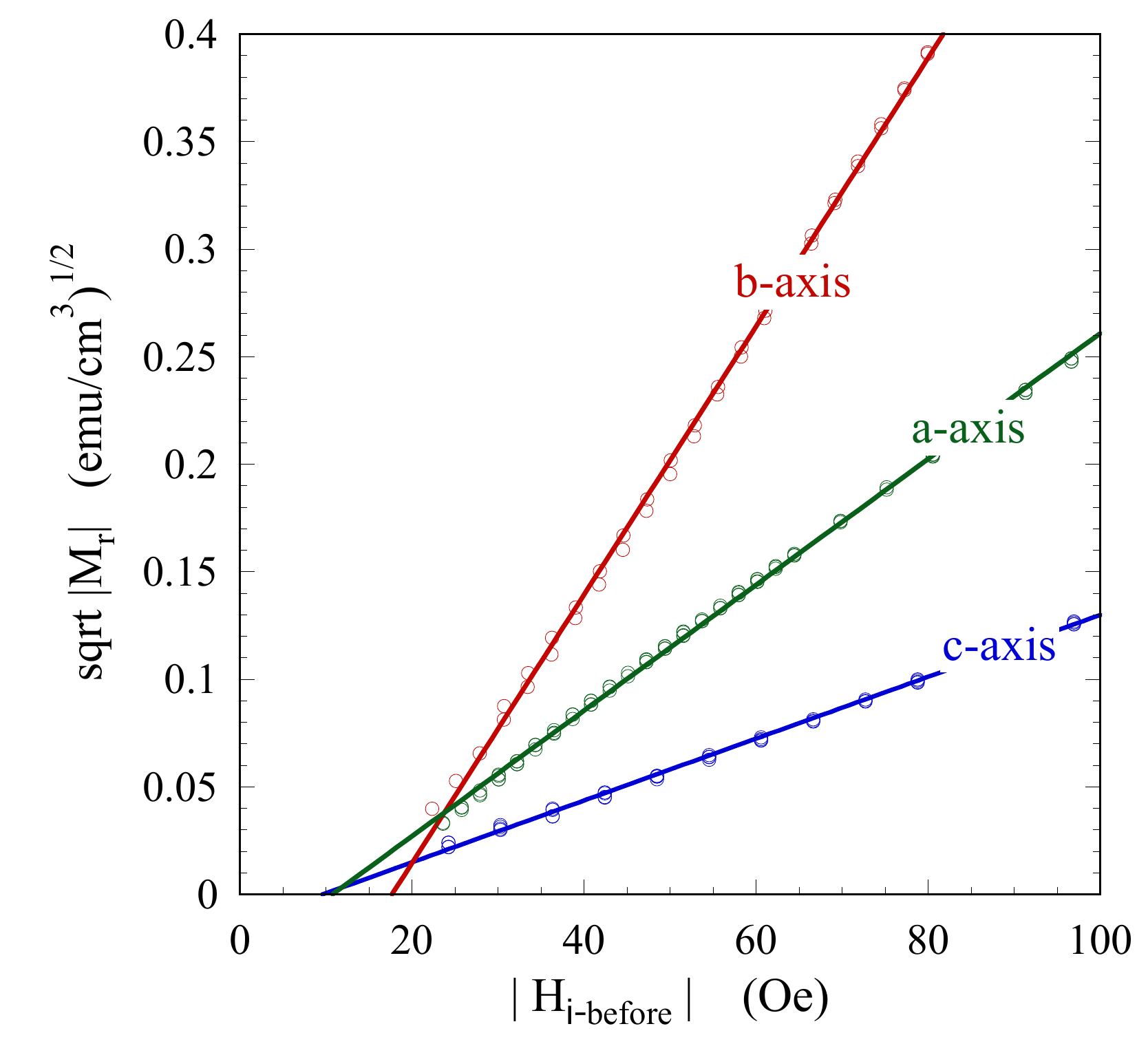} 
\caption{ %The same data for the second sample as shown in figure 16
\added[id=JP]{Same data as shown in figure \protect{\ref{suppl_fig16}}} 
, but here plotted as sqrt$( |M_{\rm r}|)$ vs $H_{\rm i-before}$, where $H_{\rm i- before}$ is the internal field after correcting for demagnetization effects.  The corrections for the $a$ and $b$ axis are modest, but for the $c$ axis where the field was perpendicular to what was left of the original slab shape, the corrections more than double the apparent field.}
\label{suppl_fig18}
\end{center}
\end{figure}

The remanent magnetization for sample 2 measured along $c$ axis is shown in Figure \ref{suppl_fig16} , plotted as sqrt$( |M_{\rm r}|)$ vs $H_{\rm before}$. The dashed lines are fits to the data above the cut-off. Even for measurements perpendicular to the slab, above the cut-off the field still penetrates to a very good approximation as $(H-H_{\rm c1})^2$. However we expect demagnetization corrections to be very important for this direction. To see this we plot $M_{\rm r}$ for all three directions in Figure \ref{suppl_fig17}  as sqrt$( |M_{\rm r}|)$ vs $H_{\rm before}$, i.e. the applied field with out demagnetization corrections, and in Figure \ref{suppl_fig18}  as  sqrt$( |M_{\rm r}|)$ vs $H_{\rm i-before}$, where $H_{\rm i- before}$ is the internal field after correcting for demagnetization effects.  As can be seen in the figures,  the corrections for the $a$ and $b$ axis are modest, but for the $c$ axis where the field was perpendicular to the slab\added[id=JP]{-}shape sample, the corrections are very important, more than doubling the apparent field.

%\newpage

\begin{figure}[h]
\makeatletter
\renewcommand{\thefigure}{S\@arabic\c@figure}
\makeatother
\begin{center}
\includegraphics[width=0.75\columnwidth]{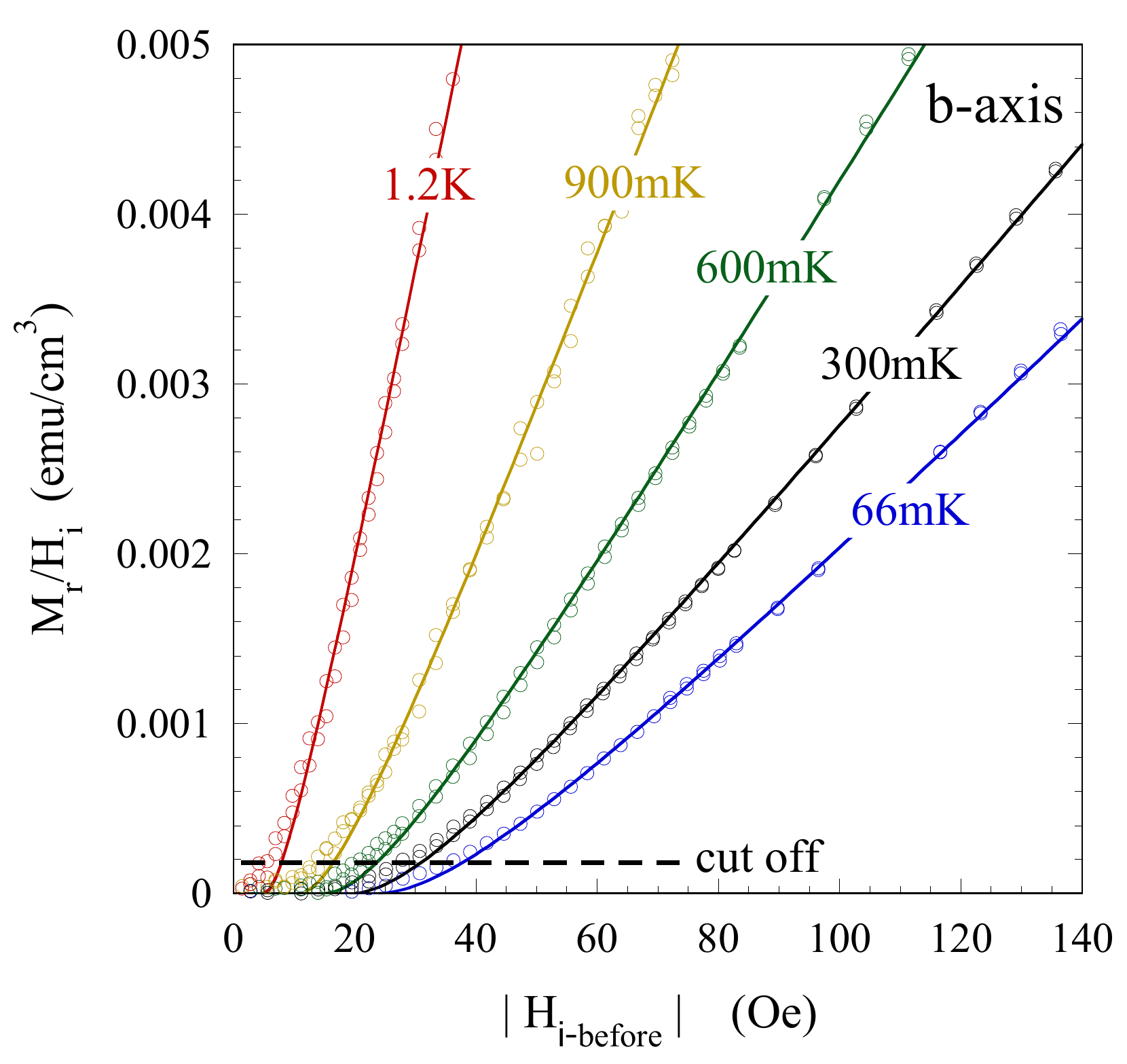} 
\caption{ The remanent magnetization for sample 2, corrected for demagnetization effects at several temperatures plotted as $M\added[id=JP]{_{\rm r}}/H_{\rm i}$ vs $H_{\rm i}$.
The solid lines are fits \replaced[id=JP]{of $M_{\rm r}/H_{\rm i}$ against}{to the data using}  $k(H_{\rm i}-H_{\rm c1})^2/H_{\rm i}$, w\added[id=JP]{h}ere $k=1/4H^*$ is inversely proportional to the critical current. 
\added[id=JP]{Hence the values of the critical current deduced from $M_{\rm r}$ reported in figure 4 of the main text.}
}
\label{suppl_fig19}
\end{center}
\end{figure}

The two parameters used to fit the data are $H_{\rm c1}$ and $1/2H^*$ for the magnetization or $H_{\rm c1}$ and $1/4H^*$ when using $M_{\rm r}$.  $1/2H^*$ and $1/4H^*$  (roughly the slope of the points at high field) \replaced[id=JP]{are}{is} inversely proportional to the critical current  $J_{\rm c}$. From the figure it can be seen that the remanent slope is approximately 1/2 the slope found from the magnetization, as it should in accordance to the critical state model outlined above.  Figure \ref{suppl_fig19}  shows examples of remanent magnetization for the second sample along the $b$ axis at several temperatures plotted as $M\added[id=JP]{_{\rm r}}/H_{\rm i}$ vs $H_{\rm i}$.  In this way we have use\added[id=JP]{d} the values of the fit  \replaced[id=JP]{of}{to} the remanent magnetization to calculate  $J_{\rm c}$ as a function of temperature, and \added[id=JP]{this} is report\added[id=JP]{ed} in figure 4 of the main text along with values obtained from the ac susceptibility.

%\clearpage

\section{ac susceptibility $\chi$ and current density $J_{\rm c}$}
\begin{figure}[H]
\makeatletter
\renewcommand{\thefigure}{S\@arabic\c@figure}
\makeatother
\begin{center}
\includegraphics[width=0.8\columnwidth]{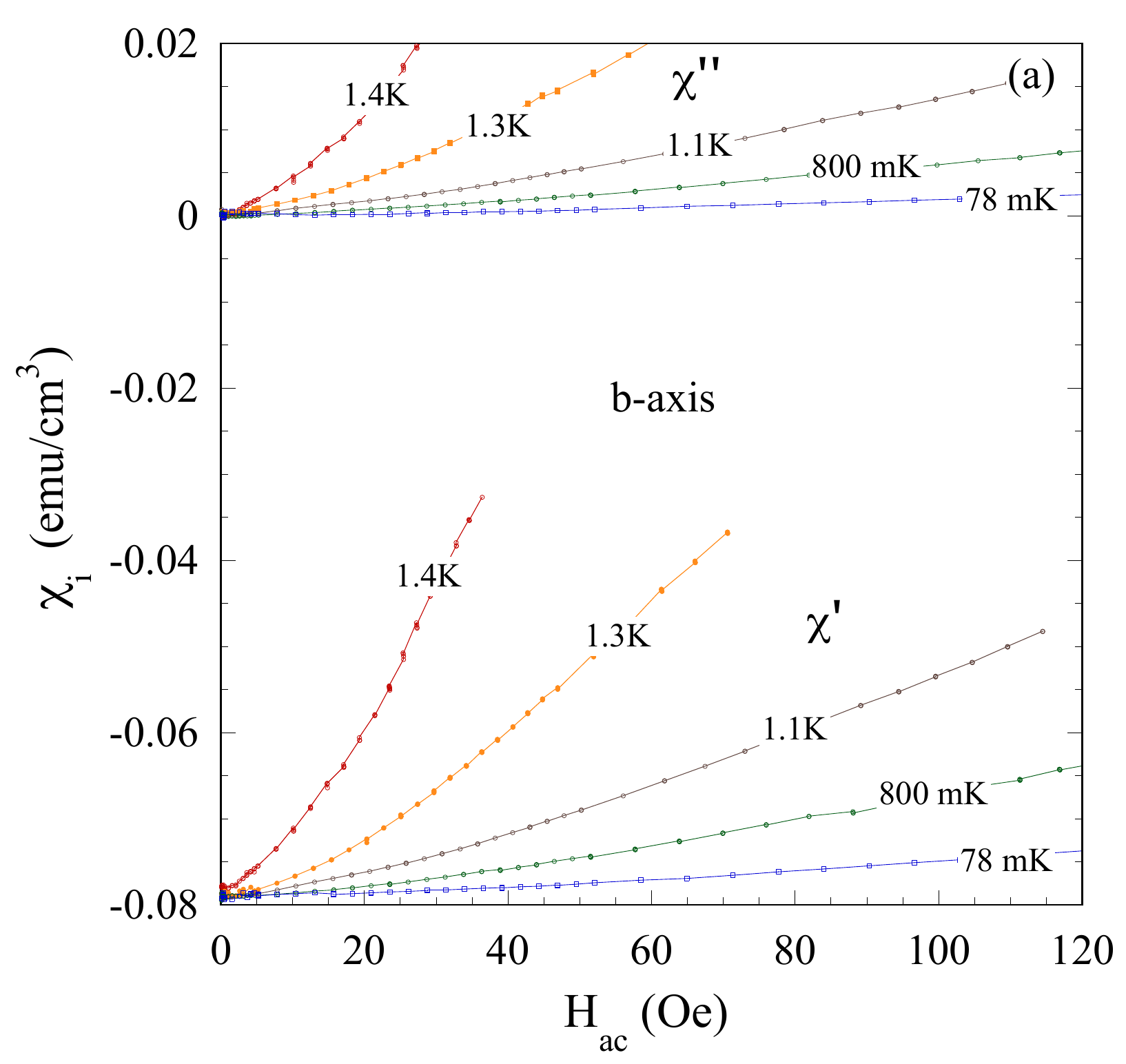} 
\caption{ The real and imaginary parts of the intrinsic ac susceptibility at 1.1~Hz vs the ac driving field $H_{\rm ac}$ for UTe$_2$ measured along the $b$ axis. The data have been corrected for demagnetization effects. When flux begins to enter the sample, $\chi'$ and $\chi''$ of the ac susceptibility will deviate from their 100\% shielding values. The deviations are linear in the applied driving field and the slopes are proportional to $2/(J_{\rm c}D)$ for $\chi'$, and $2/(3\pi J_{\rm c} D)$ for $\chi''$, where $J_{\rm c}$ is the current density in the critical state model, and $D$ is the sample width where we approximate the sample shapes as slabs.\cite{Gomory1997}
}
\label{suppl_fig20}
\end{center}
\end{figure}

\section{Upper critical field $H_{\rm c2}$}
\begin{figure}[H]
\makeatletter
\renewcommand{\thefigure}{S\@arabic\c@figure}
\makeatother
\begin{center}
\includegraphics[width=0.8\columnwidth]{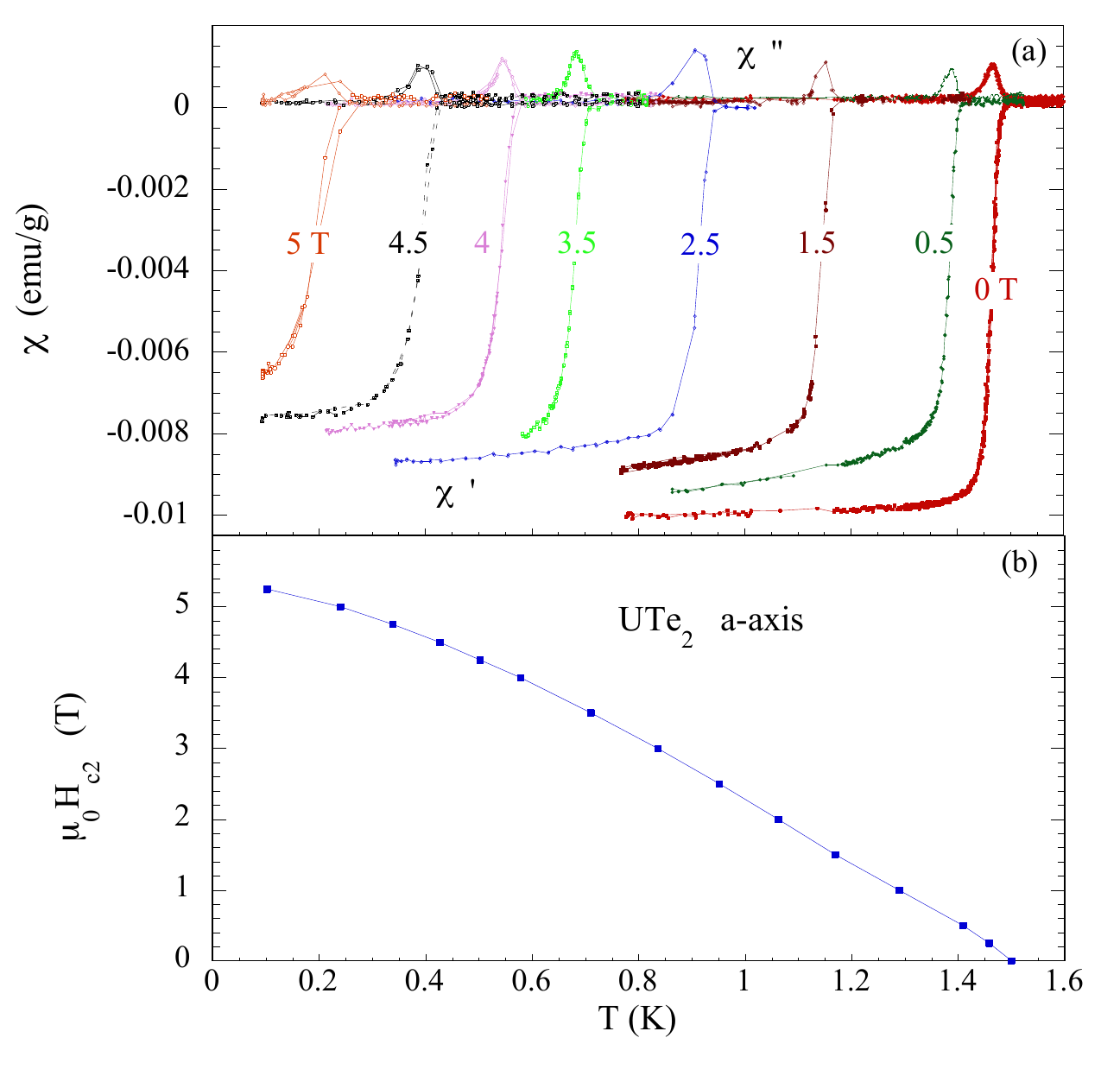} %{supplemental_figures/fig5.png}
\caption{(a) Real and imaginary parts of the susceptibility vs temperature with various dc fields ranging from 0 to 5~T. The ac driving field was 2 Oe rms at 5.7 Hz. The transition for zero dc field is sharp at 1.5~K, and shifts to lower temperatures as the dc field is increased. Just below the transition a peak in the imaginary part of the susceptibility is observed. (b) The upper critical field $H_{\rm c2}$ defined by the ac susceptibility vs field is shown as a function of temperature. }
\label{suppl_fig21}
\end{center}
\end{figure}

%\newpage
%%%%%%%%%%%%%% part Jean-Pascal 

\section{Analysis of the anisotropy of \Hcl and \Hcu }
In the main paper, we explain that a field\added[id=JP]{-}dependent pairing can influence the slopes of both \Hcl and \Hcu at \Tsc, and that this change can be expressed  from the induced derivative $\frac{dT_{\rm sc}}{dH}$ : 
\begin {equation}
\frac{dH_{\rm ci}}{dT}= \
	\frac{ \left(  \frac{ \partial H_{\rm ci}} {\partial T} \right)_{\rm H} }{1+\frac{ dT_{\rm sc}} {dH} \left( \frac{\partial H_{\rm ci}} {\partial T} \right)_{\rm H}  }
\label{Equ.SlopeHc}
\end{equation}

However, the correction factor in the denominator will be negligibly small for \Hcl, as it is 3 to 4 orders of magnitude smaller than \Hcu. 
Hence, the slope of \Hcl at \Tsc is given by the usual expressions, and can be used together with the critical thermodynamic field to estimate, for each crystallographic direction the bare GL parameter $\kappa$ \added[id=JP]{(see expression \ref{Equ.GL}). From }
 the same GL relations, $\left( \frac{\partial H_{\rm c2}} {\partial T} \right)_{\rm H}$ \added[id=JP]{can then be obtained}.

\begin{equation}
\begin{aligned}
%\kappa&=\frac{\lambda_L}{\sqrt{2}\xi}&\\
H_{\rm c1} &= \frac{H_{\rm c}}{\sqrt{2}\kappa}\left(ln(\kappa)+0.49\right) \\
H_{\rm c2} &= H_{\rm c}\sqrt{2}\kappa \\
\label{Equ.GL}
\end{aligned}
\end{equation}

Knowing $\left( \frac{\partial H_{\rm c2}} {\partial T} \right)_{\rm H}$, we can invert (\ref{Equ.SlopeHc}) to obtain the required values of $\frac{dT_{\rm sc}}{dH}$ that could explain the difference with the measured values of \Hcu. Table \ref{Table.HcTrial} reports these values. \added[id=JP2]{Please note that in table \ref{Table.HcTrial} as well as in Tab.~\ref{Table.HcDef}, the Ginzburg-Landau parameters as well as the coherence length and penetration depth have been deduced from the slopes of the critical fields at \Tsc, not from their zero-field values. Indeed, even though the Ginzburg-Landau equations are valid only close to \Tsc, it is more convenient and hence the custom to deduce these values from expressions related to their zero temperature values. This is perfectly fine for single bands weak-coupling superconductors, but may become wrong for multigap, strong coupling unconventional superconductors, which is certainly the case in \UTe.}
 
\begin{table}[h]
%%%%%%%%%%%%%

\begin{ruledtabular}
\caption{The first two lines are experimental data for the slopes of the critical fields at \Tsc. The slope of the thermodynamic critical field at \Tsc is $-5.7 \cdot 10^{-2}$~T/K. The GL parameter $\kappa$ has been deduced from \Hcl and \Hc accordind to (\ref{Equ.GL}), and then $\left( \frac{\partial H_{\rm c2}}{\partial T}\right)_{\rm H}$~(T/K) from these values and the same equations. The required field dependence of the critical temperature, (arising from the field dependent pairing), so as to match $\left( \frac{\partial H_{\rm c2}}{\partial T}\right)_{\rm H}$ with experimental data for $\left (\frac{dH_{\rm c2}}{dT} \right)$is deduced from (\ref{Equ.SlopeHc}).}
\label{Table.HcTrial}
\begin{tabular}{ccccc}
%\hline\noalign{\smallskip}
 						& $H \parallel a$ 	&	$H \parallel b$		&	$H \parallel c$	\\
%\hline
\hline\noalign{\smallskip}
$\frac{dH_{\rm c1}}{dT}$~(T/K) 	& 	$-1.13 \cdot 10^{-3}$ 	& 		$-3.3 \cdot 10^{-3}$ 	&	$-3.8 \cdot 10^{-3}$	\\	
$\frac{dH_{\rm c2}}{dT}$~(T/K) 	& 	-6.6 			& 		-20 / -35		&		-6.6		\\
$\kappa$ from \Hcl \& \Hc		& 	200 			& 		55			&		45		 \\
$\left( \frac{\partial H_{\rm c2}}{\partial T}\right)_{\rm H} $~(T/K) \
						& 	-16 			& 		-4.4			&		-3.6		 \\
$\frac{ dT_{\rm sc}} {dH}$~(K/T)&	-0.089		& 		0.2			&		0.125	\\	
%\hline\noalign{\smallskip}
\end{tabular}
\end{ruledtabular}

%%%%%%%%%%%%%
\end{table}

 However, as explained in the main text, the absolute values of \Hcl are strongly dependent on the criteria chosen to determine the first flux penetration, and they are probably less reliable than the anisotropy between the different directions. 
 So we also did the same calculation, but renormalizing all \Hcl values so \replaced[id=JP]{as to obtain }{that we obtained} a vanishing $\frac{dT_{\rm sc}}{dH}$ at \Tsc for $H \parallel c$. 
 Indeed, physically, it is expected that the $c$ axis is more or less "inert", or at least, much less sensitive to field dependent pairing as the easy $a$ axis, where strong suppression might occur like in ferromagnetic superconductors \cite{Wu2017,Nakamura2017}, or as the $b$ axis where the metamagnetic transition at $H_{\rm m}\approx 35$~T induces strong effective mass renormalizations \cite{Miyake2019, Imajo2019}. 
 In order to obtain a vanishing $\frac{dT_{\rm sc}}{dH}$ for $H \parallel c$, we need that $\kappa(H \parallel c)$ matches the value given by the measured \Hcu in this direction, hence $\kappa(H \parallel c) = 82$. The required correction factor on \Hcl is 0.63, leading to new values for $\kappa$ and $\frac{ dT_{\rm sc}} {dH}$ reported in table~\ref{Table.HcDef}. 
 We have also put in the table  the characteristic length scales of the superconducting state (coherence length $\xi_0$ and London penetration depth $\lambda_L$), deduced from the zero field Fermi velocities and from the GL parameters ($\kappa=\frac{\lambda_L}{\sqrt{2}\xi}$) of table~\ref{Table.HcDef}: 
 we give both these length along each crystallographic direction \added[id=JP2]{($ \xi_i$, $ \lambda_i$), which} depend on the average Fermi velocity in this direction, and for each field direction \added[id=JP2]{($ \xi (H \parallel i)$, $ \lambda (H \parallel i)$), remembering that} the electromagnetic response depends on the average lengths perpendicular to the field. 
 If we want to compare $\lambda_L$ to published measurements \cite{Metz2019} performed at low temperature and for very low fields $H \parallel c$: 
 $\lambda_L\approx 950-1200$~nm, then we should use, rather than the slopes, the value of $H_{\rm c1}^c(0)$ and of $\kappa$ along c, as the temperature dependence of \Hcl alo\added[id=JP]{n}g this direction is very anomalous. 
 From these data, we extract $\lambda_L \approx 1300$~nm, a fairly reasonable agreement.

 \begin{table}[h]
%%%%%%%%%%%%%
\begin{ruledtabular}
\caption{The first line is experimental data for the slope of \Hcl at \Tsc, renormalized by 0.63 so as to obtain no corrections on the slope of \Hcu for $H \parallel c$ (see text). All other figures where then deduced as explained in the caption of table~\protect{\ref{Table.HcTrial}}.}
\label{Table.HcDef}
\begin{tabular}{ccccc}
%\hline\noalign{\smallskip}
 						& $H \parallel a$ 	&	$H \parallel b$		&	$H \parallel c$	\\
\hline\noalign{\smallskip}
$\frac{dH_{\rm c1}}{dT}$~(T/K) 	& 	$-0.75 \cdot 10^{-3}$ 	& 		$-2.1 \cdot 10^{-3}$ 	&	$-2.4 \cdot 10^{-3}$	\\	
$\frac{dH_{\rm c2}}{dT}$~(T/K) 	& 	-6.6 			& 		-20 / -35		&		-6.6		\\
$\kappa$ from \Hcl \& \Hc		& 	336 			& 		98			&		82		 \\
$\left( \frac{\partial H_{\rm c2}}{\partial T}\right)_{\rm H} $~(T/K) \
						& 	-27 			& 		-7.9			&		-6.6		\\
$\frac{ dT_{\rm sc}} {dH}$~(K/T)&	-0.11			& 		0.1			&		0		\\	
$ \xi_i$~(nm)				&	14			&		4.2			&		3.5		\\
$ \lambda_i$~(nm)			&	490			&		1670			&		2000		\\
$ \xi (H \parallel i)$~(nm)		&	3.85			&		7.1			&		7.8		\\
$ \lambda (H \parallel i)$	~(nm)&	1830			&		990			&		905		\\
%\hline\noalign{\smallskip}
\end{tabular}
\end{ruledtabular}

%%%%%%%%%%%%%
\end{table}

The truth might lie between these two extremes, but in order to discuss "quantitatively" the consequences of this analysis, we chose to focus on the more phys\added[id=JP]{i}cal figures of table~\ref{Table.HcDef}. In any case, it is naturally required that the anisotropy of \Hcu, if it was controlled only by the zero field values of \Tsc and the Fermi velocities would be opposite to those measured experimentally, with  $H_{\rm c2}^b$ slightly smaller than $H_{\rm c2}^c$, and a very large $H_{\rm c2}^a$ (of order the measured values for $H_{\rm c2}^b$). This would arise from the field dependent pairing, and using the same model as for the ferromagnetic superconductors \cite{Wu2017} and already applied to \UTe \cite{Knebel2019,Knebel2020}, we can calculate how the strong coupling parameter $\lambda$ should vary in the three directions in order to reproduce the measured values of \Hcu \added[id=JP]{reported} in \cite{Knebel2019,Knebel2020}, starting from the zero field parameters of table~\ref{Table.HcDef}
 
\begin{figure}[h]
%\makeatletter
%\renewcommand{\thefigure}{S\@arabic\c@figure}
%\makeatother
\begin{center}
\includegraphics[width=0.8\columnwidth]{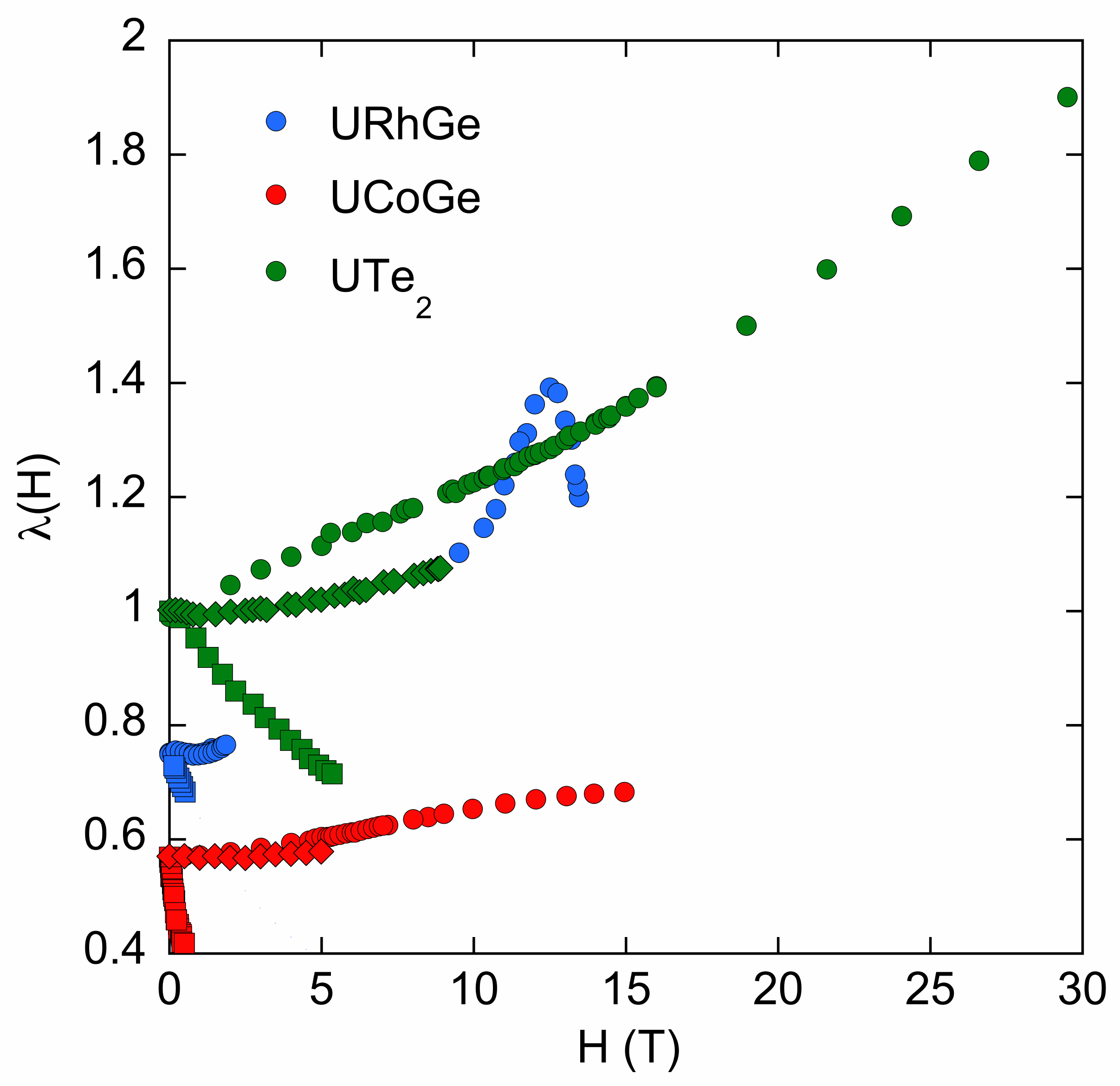}
\caption{Comparison of the field dependence of the strong coupling constant $\lambda$ in the different directions for URhGe (blue), UCoGe (red) as published in \protect {\cite{Wu2017}} and \UTe (green) as deduced from \Hcu data \cite{Knebel2019,Knebel2020} and from the zero field Fermi velocities, yielding the deduced values of $\left( \frac{\partial H_{\rm c2}}{\partial T}\right)_{\rm H}$ reported in table~\ref{Table.HcDef}. \added[id=JP]{Squares: easy axis, circles: hard axis, diamonds: intermediate axis}}
\label{Fig.lambda}

\end{center}
\end{figure}

 This is shown on Fig.\ref{Fig.lambda}, together with previous results on UCoGe and URhGe \cite{Wu2017}. Hence, the astonishing conclusion that it is the field dependence of the pairing which reverses the anisotropy of \Hcu between $a$ and $b$ axis, is less surprising when translated in terms of field dependence of the strong coupling parameter $\lambda$: this field dependence, and notably the strong suppression of $\lambda$ along the easy axis, is similar to what we expect in URhGe and much weaker than what we observe in UCoGe. Only the strong increase of  $\lambda$ for $H \parallel$ $b$ axis is very different in \UTe, but this is also consistent with the specific heat ($C_{\rm p}$) measurements \cite{Imajo2019,Miyake2019}, which do show a strong increase of  ($C_{\rm p}/T$) under field with a finite slope at zero field.

\newpage
\bibliographystyle{apsrev4-2}	
\bibliography{UTe2supp}